\documentclass[12pt]{article}
\usepackage{amsmath,amssymb}

\usepackage{typearea}
\typearea{13}

  \makeatletter
    
    \@addtoreset{equation}{section}
  \makeatother
   
\begin{document}
\renewcommand{\include}[1]{}
\renewcommand\documentclass[2][]{}

\newcommand{\ba}{\begin{alignat}{3}}

\newcommand{\e}{\epsilon}
\newcommand{\oep}{\overline{\epsilon}}
\newcommand{\dl}{\delta}
\newcommand{\lam}{\lambda}
\newcommand{\olam}{\overline{\lambda}}
\newcommand{\si}{\sigma}
\newcommand{\g}{\gamma}
\newcommand{\G}{\Gamma}
\newcommand{\tr}{\text{Tr}}
\newcommand{\w}{\wedge}
\newcommand{\pa}{\partial}
\newcommand{\tf}{\tfrac}
\newcommand{\Om}{\Omega}
\newcommand{\om}{\omega}
\newcommand{\bra}{\langle}
\newcommand{\ket}{\rangle}
\newcommand{\Ob}{\mathcal{O}}
\newcommand{\D}{\mathcal{D}}
\newcommand{\Ai}{\mathcal{A}}
\newcommand{\ops}{\overline{\psi}}
\newcommand{\Lam}{\Lambda}
\newcommand{\mon}{\text{mon}}
\newcommand{\fla}{\text{flat}}

\newcommand{\La}{\mathcal{L}}
\newcommand{\Ra}{\mathcal{R}}
\newcommand{\A}{\mathcal{A}}
\newcommand{\na}{\nabla}
\newcommand{\Dg}{D \! \! \! \! /}
\newcommand{\rt}{\sqrt{t}}
\newcommand{\Vg}{V_\mu \g^\mu }

\newcommand{\hphi}{\hat{\phi}}
\newcommand{\hpsi}{\hat{\psi}}
\newcommand{\vphi}{\varphi}
\newcommand{\vt}{\vartheta}
\newcommand{\tl}{\tilde{l}}

\newcommand{\ov}{\overline{v}}
\newcommand{\oom}{\overline{\om}}

\newcommand{\Dmue}{\g_{\mu} \g_3 \Big( \frac{1}{2f} \e  \Big) }
\newcommand{\Dmuoep}{\g_{\mu} \g_3 \Big( \frac{-1}{2f} \oep  \Big) }
\newcommand{\Dnue}{\g_{\nu} \g_3 \Big( \frac{1}{2f} \e  \Big) }
\newcommand{\Dnuoep}{\g_{\nu} \g_3 \Big( \frac{-1}{2f} \oep  \Big) }


\newcommand{\Sp}{\mathbb{S}}
\newcommand{\Rp}{\mathbb{RP}}
\newcommand{\compe}{\vartheta}
\newcommand{\compt}{\varphi}
\newcommand{\compd}{y}
\newcommand{\h}{\mathfrak{h}_n}
\newcommand{\p}{\mathfrak{p}_n}
\newcommand{\tri}{\Omega}
\newcommand{\rad}{}
\newcommand*\nor[1]{%
  \ifcase #1 \or n \or \frac{n}{\rad} \fi}
\newcommand{\darn}{\nor{1}}


\newcommand{\ch}{\text{\boldmath $q$}}
\newcommand{\fl}{\text{\boldmath $f$}}


\newcommand{\SeAmu}{ - \frac{i}{2} \olam \g_{\mu} \e  }
\newcommand{\SeAnu}{- \frac{i}{2} \olam \g_{\nu} \e  }
\newcommand{\SoeAmu}{- \frac{i}{2} \oep \g_{\mu} \lam  }
\newcommand{\SoeAnu}{- \frac{i}{2} \oep \g_{\nu} \lam  }
\newcommand{\Sesi}{+ \frac{1}{2} \olam  \e   }
\newcommand{\Soesi}{+ \frac{1}{2} \oep  \lam  }
\newcommand{\Selam}{ \frac{1}{2}  \g^{\mu \nu} \e F_{\mu \nu} - D \e + i \g^{\mu} \e \D_{\mu} \si +\frac{i}{f} \si \g_3 \e   }
\newcommand{\Soelam}{0}
\newcommand{\Seolam}{ 0  }
\newcommand{\Soeolam}{\frac{1}{2}  \g^{\mu \nu} \oep F_{\mu \nu} + D \oep - i \g^{\mu} \oep \D_{\mu} \si + \frac{i}{f} \si \g_3 \oep  }
\newcommand{\SeD}{ + \frac{i}{2}  \D_{\mu} \olam \g^{\mu} \e 
+ \frac{i}{4f} \olam \g_3 \e  }
\newcommand{\SoeD}{- \frac{i}{2} \oep \g^{\mu} \D_{\mu} \lam 
+  \frac{i}{4f} \oep \g_3 \lam  }

\newcommand{\ophi}{\overline{\phi}}
\newcommand{\opsi}{\overline{\psi}}
\newcommand{\oF}{\overline{F}}
\newcommand{\Sephi}{0}
\newcommand{\Soephi}{\oep \psi }
\newcommand{\Seophi}{\e \opsi}
\newcommand{\Soeophi}{0 }
\newcommand{\Sepsi}{ i \g^\mu \e \D_\mu^A \phi + i \ch \e \si \phi - \frac{i \Delta}{f} \e \g_3 \phi }
\newcommand{\Soepsi}{\oep F}
\newcommand{\Seopsi}{\oF \e}
\newcommand{\Soeopsi}{i \g^\mu \oep \D_\mu^A \ophi + i\ch \ophi \si \oep - \frac{i \Delta}{f}  \ophi \g_3 \oep }
\newcommand{\SeF}{i \e \g^\mu \D_\mu^A \psi - i\ch \si  \e \psi  - i \ch \e \lam \phi + \frac{i(2 \Delta -1)}{2f}\e \g_3 \psi}
\newcommand{\SoeF}{0}
\newcommand{\SeoF}{0}
\newcommand{\SoeoF}{i \oep  \g^\mu \D_\mu^A \opsi - i \ch \oep \opsi \si  + i \ch \ophi  \oep   \olam - \frac{i(2\Delta -1)}{2f} \oep \g_3 \opsi }
\newcommand{\set}{\setlength\arraycolsep{2pt}} 
\newcommand{\non}{\nonumber} 
\newcommand{\seteq}{\setcounter{equation}}

\newcommand{\cor}{\color{red}}
\newcommand{\cob}{\color{blue}}
\newcommand{\coc}{\color{cyan}}
\newcommand{\com}{\color{magenta}}
\newcommand{\cog}{\color{green}}
\newcommand{\cobk}{\color{black}}
\newcommand{\cogy}{\color[gray]}
\newcommand{\corgb}{\color[rgb]}

\newcommand{\tm}{\tilde{m}}

\newcommand{\balpha}{\bar{\alpha}} 
\newcommand{\bbeta}{\bar{\beta}} 
\newcommand{\Dl}{\Delta}
\newcommand{\ve}{\varepsilon}
\newcommand{\be}{\bar{\epsilon}}
\newcommand{\he}{\hat{\epsilon}}
\newcommand{\E}{\Epsilon}
\newcommand{\teta}{\tilde{\eta}} 
\newcommand{\hg}{\hat{\gamma}}
\newcommand{\iv}{\iota_{v}} 
\newcommand{\la}{\lambda} 
\newcommand{\bla}{\bar{\lambda}}
\newcommand{\tom}{\tilde{\omega}}
\newcommand{\tOm}{\tilde{\Omega}}
\newcommand{\bvarphi}{\bar{\varphi}} 
\newcommand{\tPsi}{\tilde{\Psi}} 
\newcommand{\sg}{\sigma} 
\newcommand{\tsg}{\tilde{\sigma}}
\newcommand{\hsg}{\hat{\sigma}}
\newcommand{\Sg}{\Sigma}
\newcommand{\txi}{\tilde{\xi}} 
\newcommand{\tzeta}{\tilde{\zeta}} 

\newcommand{\Acal}{\mathcal{A}}
\newcommand{\Bcal}{\mathcal{B}}
\newcommand{\Ccal}{\mathcal{C}}
\newcommand{\Dcal}{\mathcal{D}}
\newcommand{\Ecal}{\mathcal{E}}
\newcommand{\Fcal}{\mathcal{F}}
\newcommand{\tFcal}{\tilde{\mathcal{F}}}
\newcommand{\Gcal}{\mathcal{G}}
\newcommand{\Hcal}{\mathcal{H}}
\newcommand{\Ical}{\mathcal{I}}
\newcommand{\Jcal}{\mathcal{J}}
\newcommand{\Kcal}{\mathcal{K}}
\newcommand{\Lcal}{\mathcal{L}}
\newcommand{\Mcal}{\mathcal{M}}
\newcommand{\Ncal}{\mathcal{N}}
\newcommand{\Ocal}{\mathcal{O}}
\newcommand{\Pcal}{\mathcal{P}}
\newcommand{\Qcal}{\mathcal{Q}}
\newcommand{\Rcal}{\mathcal{R}}
\newcommand{\Scal}{\mathcal{S}}
\newcommand{\Tcal}{\mathcal{T}}
\newcommand{\Ucal}{\mathcal{U}}
\newcommand{\Vcal}{\mathcal{V}}
\newcommand{\Wcal}{\mathcal{W}}
\newcommand{\Xcal}{\mathcal{X}}
\newcommand{\Ycal}{\mathcal{Y}}
\newcommand{\Zcal}{\mathcal{Z}}

\newcommand{\gfrak}{\mathfrak{g}}

\newcommand{\lp}{\left(}
\newcommand{\rp}{\right)}
\newcommand{\lc}{\left\{}
\newcommand{\rc}{\right\}}
\newcommand{\lb}{\left[}
\newcommand{\rb}{\right]}
\newcommand{\lra}{\leftrightarrow}
\newcommand{\ra}{\rightarrow}
\newcommand{\longra}{\longrightarrow}
\newcommand{\longla}{\longleftarrow}
\newcommand{\LRA}{\Leftrightarrow}
\newcommand{\RA}{\Rightarrow}
\newcommand{\Tr}{\text{Tr}}
\newcommand{\diag}{{\rm diag}} 
\newcommand{\ind}{{\rm ind}} 
\newcommand{\Td}{{\rm Td}} 
\newcommand{\CPii}{\mathbb{C} P^{2}} 

\newcommand{\reA}[3]{\cobk (#1 , #2, #3) \cobk}
\newcommand{\re}[1]{\cobk #1 \cobk}

\newtheorem{rem}{Remark}
%
%

\renewcommand{\thefootnote}{\fnsymbol{footnote}}
\setcounter{page}{0}
\thispagestyle{empty}

\begin{flushright}
OU-HET 820
\end{flushright}
\vspace{3cm}
\begin{center}
{\LARGE {\bf Superconformal index on $\Rp^2 \times \Sp^1$ \\ and mirror symmetry
}}

\vskip1cm
{\bf Akinori Tanaka,$^{a}$\footnote{akinori@het.phys.sci.osaka-u.ac.jp}}
{\bf Hironori Mori,$^{a}$\footnote{hiromori@het.phys.sci.osaka-u.ac.jp}}
{\bf and Takeshi Morita$^{b}$\footnote{t-morita@cr.math.sci.osaka-u.ac.jp}}

{\large 
\vskip1cm
\it $^{a}$Department of Physics, Graduate School of Science, 
\\
Osaka University, Toyonaka, Osaka 560-0043, Japan}

{\large 
\vskip.5cm
\it $^{b}$Graduate School of Information Science and Technology, 
\\
Osaka University, Toyonaka, Osaka 560-0043, Japan}
\end{center}

\vskip1cm
\begin{abstract}
We study $\Ncal = 2$ supersymmetric gauge theories on $\Rp^2 \times \Sp^1$ and compute the superconformal index by using the localization technique. We consider not only the round real projective plane $\Rp^2$
but also the squashed real projective plane $\Rp^2_b$ which turns back to $\Rp^2$ by taking a squashing parameter $b$ as $1$. In addition, we find that the result is independent of the \textit{squashing parameter} $b$. We apply our new superconformal index to check the simplest case of 3d mirror symmetry, i.e. the equivalence between the $\Ncal=2$ SQED and the XYZ model on $\Rp^2 \times \Sp^1$. We prove it by using a mathematical formula called the $q$-binomial theorem. We also comment on the $\Ncal=4$ version of mirror symmetry, mirror symmetry via generalized indices, and possibilities of generalizations from mathematical viewpoints.
\end{abstract}
\renewcommand{\thefootnote}{\arabic{footnote}}
\setcounter{footnote}{0}

\vfill\eject
\tableofcontents

\section{Introduction}
The remarkable recent progress in 3d supersymmetric gauge theories is that we can exactly investigate theories with interactions on various curved geometries by making use of the localization \cite{Pestun:2007rz, Kapustin:2009kz, Gang:2009wy, Jafferis:2010un, Hama:2011ea, Kallen:2011ny, Imamura:2011wg, Ohta:2012ev, Alday:2013lba}. One of the interesting quantities to which we can apply this exact calculation is the superconformal index (SCI) \cite{Kinney:2005ej, Bhattacharya:2008zy} defined as a refinement of the Witten index. The SCI of $\Ncal = 2$ superconformal theories is defined by \cite{Bhattacharya:2008bja}
\begin{align}
\Ical ( x, e^{i \mu_{a}} ) = \Tr_{\Hcal} \lb ( - 1 )^{\hat{F}} x'^{\{ Q, Q^{\dagger} \}} x^{\hat{H} + \hat{j}_{3}} \prod_{a} e^{i \mu_{a} \hat{f}_{a}} \rb,
\label{sci}
\end{align}
where $\Hcal$ is the Hilbert space of the theory, and the trace is taken over this Hilbert space (see Section \ref{Locmain} for details). Basically, it counts the number of supersymmetric vacua, so-called BPS states, with eigenvalues of certain operators commuting with both the Hamiltonian $\{ Q, Q^{\dagger} \}$ and the fermion number operator $\hat{F}$. The SCI on $\mathbb{S}^{2} \times \mathbb{S}^{1}$ has been computed by the localization in \cite{Kim:2009wb, Imamura:2011su}.

An application of the SCI is to study 3d mirror symmetry \cite{Intriligator:1996ex, Aharony:1997bx, Kapustin:1999ha, Tong:2000ky} of which the duality between the $\Ncal = 2$ SQED and the XYZ model is the simplest example. Mirror symmetry on $\mathbb{S}^{2} \times \mathbb{S}^{1}$ based on SCIs has been studied numerically in \cite{Imamura:2011su} and has been manifested in \cite{Krattenthaler:2011da} by using the $q$-binomial theorem and Ramanujan's summation (the special  case for SCIs with gauging flavor symmetries called generalized indices also has been proven in the same way \cite{Kapustin:2011jm}). An advantage in utilizing the SCI is that we can establish mirror symmetry rigorously in the mathematical sense thanks to the localization.

On the other hand, one can construct 2d theories on the real projective plane $\Rp^2$ by taking precise boundary conditions of fields on the two-sphere $\Sp^2$ under the antipodal identification
\begin{align}
(\pi - \compe, \pi +\compt)
\sim
(\compe, \compt),
\label{antipodal2}
\end{align}
where $\compe, \compt$ are coordinates of $\Sp^2$. The partition function on $\Rp^2$ has been computed exactly in \cite{Kim:2013ola}. The authors also showed how to define 2d supersymmetry (SUSY) theories on the \textit{squashed real projective plane} $\Rp_b^2$ by turning on an appropriate background U$(1)_R$-gauge field. This method was developed in \cite{Gomis:2012wy} in the context of localization calculus on the \textit{squashed two-sphere} $\Sp^2_b$.

In this paper, we show that their constructions can be lifted naturally to these on $\Rp_b^2 \times \Sp^1$ by adding the third coordinate $\compd$. We can get this curved space from $\Sp_b^2 \times \Sp^1$ by identifying
\begin{align}
(\pi - \compe, \pi +\compt, \compd)
\sim
(\compe, \compt, \compd),
\label{antipodal3}
\end{align}
where $\compe \in [0, \pi], \compt \in[0,2\pi]$, and $\compd \in [0,2 \pi ]$. Note that there is no difference between $\Sp_b^2 \times \Sp^1$ and $\Rp_b^2 \times \Sp^1$ if we only discuss the \textit{local} quantities. The difference between them comes from the global distinction of topology and the boundary conditions of fields under the antipodal identification \eqref{antipodal3}. With these setups, we calculate the SCI of $\Ncal = 2$ supersymmetric gauge theories on $\mathbb{RP}_{b}^{2} \times \mathbb{S}^{1}$ by the localization. First of all, we take the Kaluza-Klein (KK) expansion for all fields along the $\mathbb{S}^{1}$ direction, which reduces Lagrangians on $\mathbb{RP}_{b}^{2} \times \mathbb{S}^{1}$ to the sum of Lagrangians on $\mathbb{RP}_{b}^{2}$ over KK modes. Then, the one-loop determinant of the vector multiplet and the matter multiplet can be obtained as the product of one-loop determinants on $\mathbb{RP}_{b}^{2}$ computed in \cite{Kim:2013ola}. Furthermore, we specify parity conditions named the B-parity condition, in order to make all fields consistent with the antipodal identification. The B-parity condition is concluded by plausible requirements from physical consideration. The one-loop determinant is expressed by the contribution of the $Z_{2}$-holonomy even or odd sector due to the B-parity condition. As a result, the SCI is written as the sum of each contribution when the vector multiplet is considered. This is different from the case where the SCI on $\mathbb{S}_{b}^{2} \times \mathbb{S}^{1}$ receives the contribution of the monopole as the infinite sum over integers. In addition, the one-loop determinants and the SCI on $\mathbb{RP}_{b}^{2} \times \mathbb{S}^{1}$ are independent of the squashing parameter $b$.

With our exact results, we check $\Ncal = 2$ Abelian mirror symmetry on $\mathbb{RP}_{b}^{2} \times \mathbb{S}^{1}$. Again, the B-parity condition carries a crucial role to establish this duality. We verify it exactly as the equality of the SCIs involving the $q$-shifted factorial and the basic hypergeometric series.\footnote{We follow these names used in \cite{GR, Krattenthaler:2011da} The authors of \cite{Kapustin:2011jm} use the $q$-product instead of the $q$-shifted factorial.}.
In this paper, we do not discuss the non-Abelian case because there is a slight difficulty in the classical configuration of the gauge field. Its solutions of saddle point equations consistent with the B-parity are written by the flat connection on $\mathbb{RP}_{b}^{2}$ and the Wilson line phase along $\mathbb{S}^{1}$ (see \eqref{locus}). After the localization, the final form of the index becomes the integration over the saddle points (i.e., Coulomb moduli). Besides the one-loop determinants, we have the Jacobian coming from fixing its integration measure onto the Cartan subalgebla by using the gauge symmetry of the saddle points. However, this factor is now undetermined since we do not find the explicit form of the flat connection. It is straightforward to extend all other arguments to non-Abelian SUSY gauge theories. We put off this issue as a future work.

The rest of this paper is organized as follows:
In Section \ref{SUSYbasis}, we construct $\Ncal = 2$ supersymmetric gauge theories with the U$(1)$ gauge group on $\mathbb{RP}_{b}^{2} \times \mathbb{S}^{1}$. Also, we indicate the B-parity condition for a single flavor and its generalization to $N_{f}$ flavors.
In Section \ref{Locmain}, we show the main idea of the localization computation on $\mathbb{RP}_{b}^{2} \times \mathbb{S}^{1}$ and the one-loop determinant for the vector multiplet and the matter multiplet. In addition, we provide the general formula of the new SCI for convenience.
In Section \ref{AMS}, mirror symmetry on $\mathbb{RP}_{b}^{2} \times \mathbb{S}^{1}$ is established as the relation of the SCI for the SQED and the XYZ model with the appropriate identifications of variables. We must take account of the B-parity condition correctly to get these SCIs. We justify it mathematically by employing the $q$-binomial theorem.
In Section \ref{Discussion}, we summarize our results and comment on some open questions.
In Appendixes \ref{1-loop} and \ref{Formula}, we explain calculation details of the one-loop determinants.
In addition to mirror symmetry, there is another example called an Abelian duality hold between a gauge field and a scalar in a 3d Abelian gauge theory \cite{Prodanov:1999jy, Broda:2002wg, Beasley:2014ila} to be able to check the validity of our new SCI. We show how the SCI on $\mathbb{RP}_{b}^{2} \times \mathbb{S}^{1}$ precisely works on Abelian duality in Appendix \ref{AD}.
In Appendix \ref{gene}, we discuss mathematical generalizations of our relation obtained as mirror symmetry.

\section{Supersymmetry on $\Rp_b^2 \times \Sp^1$}\label{SUSYbasis}
In this section, we show how to define 3d SUSY theories with U$(1)$ gauge group on $\Rp_b^2 \times \Sp^1$. Of course, we can also define the theories on $\Sp^2_b \times \Sp^1$. In fact, the arguments in Section \ref{Conventions}, \ref{SUSYonRP2S1}, and \ref{LagonRP2S1} can be applied to the theories on $\Sp^2_b \times \Sp^1$. However, we omit explanations for the calculations of the index on $\Sp^2_b \times \Sp^1$ because the final results are free from the squashing parameter and just reproduce the known results on $\Sp^2 \times \Sp^1$ \cite{Kim:2009wb, Gang:2009wy, Imamura:2011su}. On the other hand, discussions on $\Rp_b^2 \times \Sp^1$ produce truly new results even though they are free from the squashing parameter. Therefore, we concentrate on the explanations of the theories on $\Rp_b^2 \times \Sp^1$.

\subsection{Our background and conventions}\label{Conventions}
We extend the construction of 2d Killing spinors and the background U$(1)_R$-gauge field in \cite{Gomis:2012wy, Kim:2013ola} to the 3d case.

\paragraph{Our background}
We consider the following dreibein and background U$(1)_R$-gauge field:
\begin{align}
&
e^1 = f(\compe) d \compe,
\quad
e^2 = l \sin \compe d \compt,
\quad
e^3 = d\compd, 
\label{dreibein}
\\
&
V^R = 
\frac{1}{2} (1 -\frac{l}{f} ) d \compt
+
\frac{-i}{2l} (1 -\frac{l}{f} ) d\compd
,
\label{VR3}
\end{align}
where $f(\compe)^2=\tilde{l} \ ^2 \sin^2 \compe+l^2 \cos^2 \compe$.  We use latin alphabet $a, b, c$ for the local Lorentz indices.

\paragraph{Covariant derivative}
The 3d covariant derivative is defined by
\begin{align}
\D_\mu = \pa_\mu + \frac{1}{4} \om^{ab}_\mu \hat{\mathcal{J}_{ab} }-i \hat{R} V^R_\mu,
\label{cov3}
\end{align}
where $\om_\mu^{ab}$ is the spin connection computed from the dreibein \eqref{dreibein}, and the $\hat{\mathcal{J}_{ab}}$'s are Lorentz generators of the fields characterized by its spin:
\begin{align}
\left. \begin{array}{ll}
\text{spin-0} &\Rightarrow \hat{\mathcal{J}_{ab} }=0,  \\
\text{spin-1/2} &\Rightarrow \hat{\mathcal{J}_{ab} }= \g_{ab},
\end{array} \right.
\end{align}
where $\g_{ab}$ are antisymmetrized gamma matrices defined in \eqref{gamma}, and $\hat{R}$ is an R-charge. See Table \ref{tab:A1} for  assignments of R-charges to each field.
\begin{table}[tbp]
\caption{Charge assignments for each field
\label{tab:A1}}
\begin{tabular}{|c||c|c|} 
\hline
Killing spinor & $\e$ & $\overline{\e}$\\ \hline \hline
Spin & 1/2 & 1/2  \\ \hline
$\hat{R}$ & +1 & $-1$  \\ \hline
\end{tabular}
\centering
\begin{tabular}{|c||c|c|c|c|c||c|c|c|c|c|c|} 
\hline
Field & $A_{\mu}$ & $\si$ & $\lam$ & $\olam$ & $D$  & $\phi$ & $\overline{\phi}$ & $\psi$ & $\overline{\psi}$ & $F$ & $\overline{F}$ \\ \hline \hline
Spin & 1 & 0 & 1/2 & 1/2 & 0  & 0 & 0 & 1/2 & 1/2 & 0 & 0 \\ \hline
$\hat{R}$ & 0 & 0 & +1 & $-1$ & 0 & $-\Delta$ & $\Delta$ & $-(\Delta - 1)$ & $\Delta-1$ & $-(\Delta-2)$ & $\Delta-2$ \\ \hline
\end{tabular}
\end{table}

\paragraph{Killing spinors}
Then, the spinors
\begin{align}
\e (\compe, \compt, \compd) =
e^{\frac{1}{2} (\frac{\compd}{l} + i \compt)}
\begin{pmatrix}
\cos \frac{\compe}{2} \\
\sin \frac{\compe}{2}
\end{pmatrix}
,\quad
\oep (\compe, \compt, \compd) =
e^{\frac{-1}{2} (\frac{\compd}{l} + i \compt)}
\begin{pmatrix}
\sin \frac{\compe}{2} \\
\cos \frac{\compe}{2}
\end{pmatrix}
\label{kil3}
\end{align}
satisfy the Killing spinor equations,
\begin{align}
\D_\mu \e=
\frac{1}{2f} \g_\mu \g_3 \e
,\quad
\D_\mu \oep=
\frac{-1}{2f} \g_\mu \g_3 \oep
.
\label{sqkse3}
\end{align}

\paragraph{Gamma matrices}
The gamma matrices $\g_a$ are defined by the Pauli matrices
\begin{align}
\g_1 = \begin{pmatrix}
0 & 1 \\
1 & 0
\end{pmatrix}
,
\quad
\g_2 = \begin{pmatrix}
0 & -i \\
i & 0
\end{pmatrix}
,
\quad
\g_3 = \begin{pmatrix}
1 & 0 \\
0 & -1
\end{pmatrix},
\quad
\g_{ab} =
\frac{1}{2}(\g_a \g_b - \g_b \g_a).
\label{gamma}
\end{align}

\paragraph{Spinor bilinear}
Let us denote generic spinors by $\e, \oep$, and $\lam$. We take spinor bilinears as
\begin{align}
&\e \lam
=
\begin{pmatrix}
\e_1 & \e_2
\end{pmatrix}
\begin{pmatrix}
0 & 1\\
-1 & 0
\end{pmatrix}
\begin{pmatrix}
\lam_1 \\ \lam_2
\end{pmatrix},
\quad
\e \g_a \lam
=
\begin{pmatrix}
\e_1 & \e_2
\end{pmatrix}
\begin{pmatrix}
0 & 1\\
-1 & 0
\end{pmatrix}
\g_a
\begin{pmatrix}
\lam_1 \\ \lam_2
\end{pmatrix}.
\notag
\end{align}
Using this convention, one can prove the following formulas:
\begin{align}
\left. \begin{array}{ll}
\e \lam = (-1)^{1+|\e|\cdot |\lam|} \lam \e,
\quad
\e \g_a \lam = (-1)^{|\e|\cdot |\lam|} \lam \g_a \e,
\quad
(\g_a \e) \lam = - \e \g_a \lam,
\\
\oep (\e \lam) + (-1)^{1+|\e|\cdot|\oep|} \e (\oep \lam) + (\oep \e) \lam =0,
\quad
 (-1)^{1+|\e|\cdot|\oep|}\e (\oep \lam) + 2 (\oep \e) \lam + (-1)^{1+|\lam|\cdot|\e|}(\oep \g_a \lam) \g^a \e =0,
\end{array} \right.
\notag
\end{align}
where $|\e|$ means the spinor $\e$'s statistics such that $|\e| = 0$ for a bosonic $\e$ and $|\e|=1$ for a fermonic $\e$.

\subsection{Supersymmetry}\label{SUSYonRP2S1}
We show our definition of supersymmetry in this subsection. There are two kinds of multiplets in the 3d $\mathcal{N}=2$ theory. The first one is the vector multiplet composed of gauge field $A_\mu$, scalar $\si$, gauginos $\lam ,\olam$, and an auxiliary field $D$. The supersymmetry for the vector multiplet is given by
\begin{align}
&\dl_\e A_\mu = \SeAmu,
\quad
\dl_{\oep} A_\mu = \SoeAmu,
\label{dA}
\\
&\dl_\e \si = \Sesi,
\quad
\dl_{\oep} \si = \Soesi,
\label{dsi}
\\
& \dl_\e \lam = \Selam,
\quad
\dl_{\oep} \lam = \Soelam,
\label{dlam}
\\
& \dl_\e \olam = \Seolam,
\quad
\dl_{\oep} \olam = \Soeolam,
\label{dolam}
\\
& \dl_\e D = \SeD,
\quad
\dl_{\oep} D = \SoeD,
\label{dD}
\end{align}
where we use the same supersymmetry as in \cite{Tanaka:2013dca}, which takes $\dl_{\e}$ and $\dl_{\oep}$ to be purely fermionic. We use the Killing spinors in \eqref{kil3} as supersymmetry parameters. $f$ is the function that appeared in \eqref{dreibein}. One can verify that the above SUSY closes within the translation, rotation, R-symmetry, and the gauge transformation.

The second one is the matter multiplet composed of scalars $\phi, \ophi$, spinors $\psi, \opsi$, and auxiliary fields $F, \oF$ which couple to the vector multiplet via the gauge symmetry with a charge $\ch$. Also, we have the following SUSY transformations for the matter multiplet:
\begin{align}
&\dl_\e \phi = \Sephi,
\quad
\dl_{\oep} \phi = \Soephi,
\label{dphi}
 \\
&\dl_\e \ophi = \Seophi ,
\quad
 \dl_{\oep} \ophi = \Soeophi ,
 \label{dpophi}
 \\
&\dl_\e \psi = \Sepsi ,
\quad
\dl_{\oep} \psi = \Soepsi ,
\label{dpsi}
\\
&\dl_\e \opsi = \Seopsi ,
\quad
\dl_{\oep} \opsi = \Soeopsi ,
\label{dopsi}
\\
&\dl_\e F = \SeF ,
\quad
\dl_{\oep} F = \SoeF ,
\label{dF}
\\
&\dl_\e \oF = \SeoF ,
\quad
\dl_{\oep} \oF = \SoeoF.
\label{doF}
\end{align}
Of course, the SUSY algebra is closed within the translation, rotation, R-symmetry, and the gauge transformation.
Here, we use the covariant derivative coupled with $A$,
\begin{align}
\D_\mu^A \Phi = \D_\mu \Phi -i \ch A_\mu \Phi,
\quad
\D_\mu^A \overline{\Phi} = \D_\mu \overline{\Phi} +i \ch \overline{\Phi} A_\mu .
\label{covderiv}
\end{align}
If one wants a neutral matter, it is achieved by turning off $\ch$. In our convention, $A_\mu$ has the same dimension as $\pa_\mu$; thus, the charge $\ch$ must be dimensionless, or, equivalently, $\ch$ is just a number. One of the most important features of the above SUSY is nilpotency
\begin{align}
\dl_\e^2 = \dl_{\oep}^2 =0.
\label{nils}
\end{align}

\subsection{Lagrangians}\label{LagonRP2S1}
\paragraph{SUSY-exact terms} 
In order to use the localization method, we need so-called \textit{SUSY-exact} terms for the vector multiplet and the matter multiplet. For the vector multiplet, we can use the super-Yang-Mills term as a SUSY-exact term. In fact, one can verify
\begin{align}
\mathcal{L}_{\rm YM}
&
=
 \frac{1}{2}    F_{\mu \nu}  F^{\mu \nu}
 + D^2
 +\D_\mu \si  \D^\mu \si
 +   \e^{3 \rho \si} \frac{\si }{f} F_{\rho \si} 
 + \frac{\si^2}{f^2}   
+ i \olam \g^\mu \D_\mu \lam
- \frac{i}{2f} \olam \g_3 \lam
\notag
\\
&=
\dl_{\oep}
\Big( ( \dl_{\oep \to \oep^{\dagger}} \olam )  \olam   \Big)
\label{YM}
\end{align}
up to a total derivative term. The notation $\dl_{\oep \to \oep^{\dagger}}$ is defined in the same way in \cite{Tanaka:2013dca}. In addition, the following term for the matter multiplet is also SUSY-exact:
\begin{align}
\mathcal{L}_{\rm mat}
&=
-i  ( \opsi \g^\mu \D_\mu^A\psi)
 +i \ch ( \opsi \si  \psi) 
- i  \ch \ophi  (\olam \psi)
- \frac{i\Delta}{2f} ( \opsi \g_3 \psi)
  +\oF F
\notag \\ & \quad
+ i  \ch(  \opsi \lam)  \phi 
+ g^{\mu \nu} \D_\mu^A \ophi \D_\nu^A \phi  
+   \ch^2 \ophi \si^2 \phi  
 + i   \ch \ophi D  \phi  \
 - \frac{\Delta}{f}    \ophi \D_3^A \phi
- \frac{\Delta}{2f^2}  \ophi \phi
+ \frac{\Delta}{4} R \ophi \phi
\notag \\
& \quad
-\frac{\Delta-1}{f}v^i \ophi  \D_i^A \phi
-
\frac{\Delta-1}{f}
\om
 \ch \ophi \si \phi
-i\frac{\Delta-1}{2f}
 v^i (\opsi \g_i \psi)
-i
\frac{\Delta-1}{2f}
 \om (\opsi \psi)
\notag
\\
&=
\dl_{\oep} \Big(
 \dl_{\oep \to \oep^{\dagger}}[\oF \phi  ]
-
i \frac{\Delta-1}{f} \dl_{\e} [\ophi \phi]
\Big),
\label{mat}
\end{align}
where $i$ runs for $1,2$, or equivalently, $\compe, \compt$. Here, we define 
\begin{align}
v^\mu = \oep \g^\mu \e 
,
\quad
\om = \oep \e.
\label{vandom}
\end{align}
We use these actions as ``regulators'' for the localization.
Thanks to the nilpotency \eqref{nils}, these terms are \textit{$\dl_{\be}$-invariant} automatically.

\paragraph{Other terms} 
Of course, we can construct other SUSY-invariant terms. The famous one is the supersymmetric Chern-Simons term
\begin{align}
\mathcal{L}_{\rm CS} =
\frac{1}{\sqrt{g}} \e ^{\mu \nu \lam}(A_{\mu} \pa_{\nu} A_{\lam}
) 
- \olam \lam + 2 D \si 
.
\label{CS}
\end{align}
This term is, however, prohibited on $\Rp^2_b \times \Sp^1$ as we will explain later. We consider the U$(1)$ gauge group; therefore, the Fayet-Iliopoulos term
\begin{align}
\mathcal{L}_{\rm FI} =
D- \frac{1}{f} A_3
\label{FI}
\end{align}
can be taken into account. If there is an additional dynamical vector multiplet, say, $(B_\mu , \tilde{\si}, \tilde{\olam}, \tilde{\lam}, \tilde{D})$, which has the same SUSY transformations as \eqref{dA} - \eqref{dD}, then we can write down the following supersymmetric BF term:
\begin{align}
\mathcal{L}_{\rm BF}
=
\frac{1}{\sqrt{g}} \e^{\mu \nu \lam} (B_\mu F_{\nu \lam})
-  \olam \tilde{\lam}
-  \tilde{\olam} \lam
+ 2 D \tilde{\si}
+ 2 \tilde{D} \si.
\label{BF}
\end{align}
However, this term is also prohibited on $\Rp^2_b \times \Sp^1$. In addition, the superpotential terms for the matter multiplet are also SUSY-invariant. It may be interesting to construct them explicitly on our curved space. For example, there is a known result how to write them explicitly on $\Sp^3$ \cite{Samsonov:2014pya}.

\subsection{Parity conditions}
As studied in \cite{Kim:2013ola}, we have to find parity conditions compatible with the antipodal identification \eqref{antipodal3} for component fields. Our guiding principles are as follows:
\begin{itemize}
 \item The squared parity transformation becomes $+1$ for bosons and $-1$ for fermions.
 \item The regulator Lagrangians \eqref{YM} and \eqref{mat} must be invariant under the parity.
 \item SUSY $\dl_\e, \dl_{\oep}$ must be consistent with the parity.
\end{itemize}

\paragraph{Vector multiplet}
After simple calculus, we find a set of parity conditions for the vector multiplet,
\begin{align}
\left. \begin{array}{ll}
A_1 (\pi - \compe, \pi + \compt, \compd) = - A_1(\compe, \compt, \compd),
\quad 
A_{2,3} (\pi - \compe, \pi + \compt, \compd) = + A_{2,3}(\compe, \compt, \compd),
 \\
\si(\pi - \compe, \pi + \compt, \compd)= -\si (\compe, \compt, \compd), \\
\lam (\pi - \compe, \pi + \compt, \compd) = +i \g_1 \lam (\compe, \compt, \compd),
\quad
\olam (\pi - \compe, \pi + \compt, \compd) = - i \g_1 \olam (\compe,  \compt, \compd), \\
D(\pi - \compe, \pi + \compt, \compd)=+D(\compe, \compt, \compd).
\end{array} \right.
\label{vecp}
\end{align}
These are similar to the ones in \cite{Kim:2013ola} called \textit{B-parity}. Therefore, we would like to call the conditions in \eqref{vecp} the B-parity condition.

\paragraph{Matter multiplet}
The one flavor matter multiplet has two choices:
\begin{align}
	\begin{aligned}
	&\phi (\pi - \compe, \pi + \compt, \compd) = \pm \phi (\compe, \compt, \compd),
	&&\ophi (\pi - \compe, \pi + \compt, \compd) = \pm \ophi (\compe, \compt, \compd), \\ 
	&\psi (\pi - \compe, \pi + \compt, \compd) = \mp i \g_1 \psi (\compe, \compt, \compd),
	&&\opsi (\pi - \compe, \pi + \compt, \compd) = \pm i \g_1 \opsi (\compe, \compt, \compd), \\ 
	&F (\pi - \compe, \pi + \compt, \compd) = \pm F(\compe, \compt, \compd),
	&&\oF (\pi - \compe, \pi + \compt, \compd) = \pm \oF (\compe, \compt, \compd). 
	\end{aligned}
\label{sematp}
\end{align}
The existence of the compatible two choices can be regarded as the existence of a holonomy with respect to a background U$(1)_{\text{flavor}}$ flat gauge field $B_\fla^{\text{flavor}}$. In other words, the parity conditions are characterized by the holonomy of $B_\fla^{\text{flavor}}$,
\begin{align}
\pm1 = e^{i \oint_\g \fl B_\fla^{\text{flavor}}},
\end{align}
where $\g$ is a noncontractible cycle on $\Rp^2$, and $\fl$ is the corresponding U$(1)_{\text{flavor}}$ charge defined by $\hat{f} \Phi = \fl \Phi$. $\hat{f}$ is a flavor charge operator used later in the definition of the superconformal index. This is an analogue of the background U$(1)_{\text{flavor}}$ monopole gauge field on $\Sp^2 \times \Sp^1$ in \cite{Kapustin:2011jm}. If we have many flavors, we can generalize these conditions. Let us denote a multiflavor field by
\begin{align}
\vec{\Phi}
=
\begin{pmatrix}
\Phi_1 \\
\Phi_2 \\
\vdots \\
\Phi_{N_f}
\end{pmatrix}
.
\end{align}
Then, the generic B-parity condition is
\begin{align}
	\begin{aligned}
	&\vec{\phi} (\pi - \compe, \pi + \compt, \compd) = {\bf M} \vec{\phi} (\compe, \compt, \compd),
	&&\vec{\ophi} (\pi - \compe, \pi + \compt, \compd) = {\bf N} \vec{\ophi} (\compe, \compt, \compd), \\ 
	&\vec{\psi} (\pi - \compe, \pi + \compt, \compd) = - i \g_1 {\bf M} \vec{\psi} (\compe, \compt, \compd),
	&&\vec{\opsi} (\pi - \compe, \pi + \compt, \compd) =  i \g_1 {\bf N} \vec{\opsi} (\compe, \compt, \compd), \\ 
	&\vec{F} (\pi - \compe, \pi + \compt, \compd) = {\bf M} \vec{F}(\compe, \compt, \compd),
	&&\vec{\oF} (\pi - \compe, \pi + \compt, \compd) = {\bf N} \vec{\oF} (\compe, \compt, \compd), 
	\end{aligned}
\label{ematp}
\end{align}
where ${\bf M}$ and ${\bf N}$ are $N_f \times N_f$ matrices constrained by
\begin{align}
{\bf N}^{\rm T} {\bf M} = {\bf 1},
\quad
{\bf M}^2 = {\bf N}^2 = {\bf 1}.
\end{align}

\paragraph{Comments on the B-parity condition} 
There are two comments. The first one is related to the vector multiplet. In order to use $\mathcal{L}_{\rm YM}$ \eqref{YM} as a regulator when we perform the localization, we would like to maintain it to be invariant under the B-parity \eqref{vecp} as we noted in our guiding principles. In fact, $\mathcal{L}_{\rm YM}$ is invariant under \eqref{vecp}. On the other hand, $\mathcal{L}_{\rm CS}$ \eqref{CS} and $\mathcal{L}_{\rm BF}$ \eqref{BF} become parity odd
\begin{align}
\mathcal{L}_{{\rm CS}/{\rm BF}} (\pi - \compe, \pi + \compt, \compd)
=
- \ \mathcal{L}_{{\rm CS}/\rm BF} (\compe , \compt, \compd).
\label{csbfparity}
\end{align}
It means that we cannot take it on $\Rp_b^2 \times \Sp^1$ as we commented just below \eqref{CS} and \eqref{BF}.

The second comment concerns the matter multiplet. Suppose we have two flavors and the B-parity condition described by the $2 \times 2$ matrices 
\begin{align}
{\bf M} = {\bf N} =
\begin{pmatrix}
0 & 1\\
1 & 0
\end{pmatrix}.
\label{XY}
\end{align}
Then, we can lift its Lagrangian on $\Rp_b^2 \times \Sp^1$ to the one on $\Sp_b^2 \times \Sp^1$ by defining a new matter multiplet on $\Sp_b^2 \times \Sp^1$ as
\begin{align}
\Phi(\compe, \compt, \compd)
=
\left\{ \begin{aligned}
& \Phi_1(\compe, \compt, \compd),
&& \compe \in [0 , \frac{\pi}{2}],
\\
& \Phi_2(\compe, \compt, \compd),
&& \compe \in [\frac{\pi}{2} , \pi].
\end{aligned} \right.
\label{doubling}
\end{align}
The authors of \cite{Kim:2013ola} also commented on this fact. This is quite similar to the \textit{doubling trick} in string theory. In Section \ref{AMS}, we use such B-parity condition exactly in the context of 3d mirror symmetry.

\section{Localization calculus on $\Rp_b^2 \times \Sp^1$}\label{Locmain}
In this section, we calculate the superconformal index (SCI)
\begin{align}
\mathcal{I}(x, \alpha)
=
\tr_{\mathcal{H}_{\Rp_b^2}}
\Big[
(-1)^{\hat{F}} x{^\prime}{^{\hat{H}+\hat{R}-\hat{j}_3} } x^{\hat{H}+\hat{j}_3} \alpha^{\hat{f}}
\Big],
\label{indexdef}
\end{align}
where $\hat{F}$ is the fermion number operator, $\hat{H}$ is the energy operator, $\hat{R}$ is the R-charge operator, $\hat{j}_3$ is the third component of the orbital angular momentum operator which acts on $\Rp_b^2$, and $\hat{f}$ is the flavor charge operator. Note that we have opposite R-charge assignments compared with \cite{Imamura:2011su, Krattenthaler:2011da, Kapustin:2011jm, Imamura:2012rq}. $\mathcal{H}_{\Rp_b^2}$ represents the Hilbert space of the theory on $\Rp_b^2$. The squashing procedure is compatible with the definition \eqref{indexdef} because this procedure preserves the isometry generated by $\hat{j}_3$. We take each fugacity as
\begin{align}
x^\prime = e^{- \beta_1},
\quad
x = e^{-\beta_2},
\quad
\alpha = e^{ i \mu},
\label{fugas}
\end{align}
where $\mu$ is a chemical potential and define the relations
\begin{align}
\beta_1 + \beta_2
= \frac{2 \pi \rad}{l},
\quad
\tri =
\frac{\beta_1 -\beta_2}{\beta_1 + \beta_2},
\label{betacond}
\end{align}
where we introduce the parameter $\tri$ for later simplicity.

\subsection{Vector multiplet contribution}
First, we have to identify the \textit{locus} of the Lagrangian $\mathcal{L}_{\rm YM}$ \eqref{YM} characterized by $\Lcal_{\rm YM} = 0$.
In order to find it, it is useful to introduce the combination of the fields
\begin{align}
\mathcal{F}^{\mu} = \frac{1}{2} \e^{\mu \rho \si } F_{\rho \si} +  \pa^\mu \si + \frac{1}{f} \dl^{\mu}_{3} \si .
\label{Fmu}
\end{align}
The Lagrangian $\mathcal{L}_{\rm YM}$ can be rewritten as
\begin{align}
\mathcal{L}_{\rm YM} =
\mathcal{F}_\mu  \mathcal{F}^{\mu}
 + D^2
+ i \olam \g^\mu \D_\mu \lam
- \frac{i}{2f} \olam \g_3 \lam
\label{YM'}
\end{align}
up to total derivative.
Now, the locus is obtained by
\begin{align}
&\mathcal{F}_\mu =0,
\label{Fmucond}
\\
&D =0,
\quad
\lam=0,
\quad
\olam=0.
\label{zeros}
\end{align}

\paragraph{Locus on $\Rp_b^2 \times \Sp^1$}
A nontrivial equation is \eqref{Fmucond}. This is equivalent to the following equation expressed by differential forms:
\begin{align}
* F 
+ d \si
+ \frac{e^3}{f} \si =0
.
\label{BPSrp2}
\end{align}
We have to know the configuration invariant under the B-parity condition \eqref{vecp} which satisfies \eqref{BPSrp2}. It can be characterized by
\begin{align}
F=0,
\quad
\si =0.
\label{BPSrp2inv}
\end{align}
The first equation in \eqref{BPSrp2inv} means, of course, the flat connection. The flat connection $\bold{A}$ on $\Rp^2 \times \Sp^1$ is expressed by
\begin{align}
\bold{A}= A_\fla + \frac{\theta}{2 \pi \rad} d \compd ,
\label{locus}
\end{align}
where $A_\fla$ is a flat connection of $\Rp^2$ related to the holonomy along the noncontractible cycle $\g$ of $\Rp^2$. There are two choices for $A_\fla = A_\fla^{(\pm)}$ characterized by
\begin{align}
\exp{\Big(i \oint_\g A_\fla^{(\pm)}  \Big)}
=\pm1,
\label{rp2fla}
\end{align}
Also, there is a constraint on $\theta$ as
\begin{align}
\theta \sim \theta + 2 \pi.
\label{theta}
\end{align}
Therefore, we have to sum up these contributions weighted by the Gaussian parts, or, equivalently, the \textit{one-loop determinants} $\Zcal_\text{1-loop}^{(\pm)}$,
\begin{align}
\mathcal{I}(x, \alpha)
=
\int_0^{2\pi} \frac{d \theta}{2 \pi}
\mathcal{Z}_\text{1-loop}^{(+)}
+
\int_0^{2\pi} \frac{d \theta}{2 \pi}
\mathcal{Z}_\text{1-loop}^{(-)}
.
\label{gaugecont}
\end{align}
One important thing is that we can perform calculus even if we do not know the explicit form of $A_\fla^{(\pm)}$. This is similar to the calculation of the partition function on $\Rp_b^2$ in \cite{Kim:2013ola}.

\paragraph{3d to 2d}
One might think that the U$(1)$ vector multiplet contribution is trivial because the result on $\Sp^2 \times \Sp^1$ was so  \cite{Kim:2009wb, Gang:2009wy, Imamura:2011su}. However, \textit{there is} a nontrivial contribution once we put the theory on $\Rp^2_b \times \Sp^1$. We can use results of 2d calculations \cite{Kim:2013ola} to compute our 3d SCI \eqref{gaugecont}. Let us show how it works. First, we expand each component field around the loci \eqref{zeros}, \eqref{BPSrp2inv}, and \eqref{locus}, then we get the following linearized Lagrangians:
\begin{align}
&\mathcal{L}_{\text{boson}}
=
 \frac{1}{2}[\pa_\mu A_\nu - \pa_\nu A_\mu]^2
  +(\pa_\mu \sigma) ^2
 +    \e^{3 \mu \nu} \frac{\si}{f} [\pa_\mu A_\nu - \pa_\nu A_\mu] 
 + \frac{\si^2}{f^2},
 \label{bos1}
 \\
&\mathcal{L}_{\text{fermion}}
=
i \olam \g^\mu \D_\mu \lam
- \frac{i}{2f} \olam \g_3 \lam
 \label{fer1}.
\end{align}
Here, our starting Lagrangian has only a U$(1)$ gauge symmetry. In other words, \eqref{YM} is the one of a Gaussian-type theory. Therefore, the above Lagrangians have nothing but the same form as the original one \eqref{YM}.

Usually, in the context of localization calculus, one expand these fields with respect to the direct product of some harmonic functions on $\Rp^2$ and Kaluza-Klein modes of $\Sp^1$.
Here, however, we take a much quicker route.
We expand each field with respect to the Kaluza-Klein modes \textit{only}:
\begin{align}
&A_i = \sum_{n }
\frac{1}{\sqrt{2\pi \rad}} e^{\big(i \darn -\frac{\beta_1 - \beta_2}{2\pi \rad}  \hat{j}_3\big)\compd }
A^{(n)}_i (\compe,\compt) 
\quad (i = 1,2),
\\
&A_3 = \sum_{n } \frac{1}{\sqrt{2\pi \rad}} e^{ \big( i \darn  -\frac{\beta_1 - \beta_2}{2\pi \rad} \hat{j}_3 \big) \compd  }
 A_3^{(n)}(\compe,\compt) ,
\\
&\si = \sum_{ n } \frac{1}{\sqrt{2\pi \rad}} e^{\big(i \darn  -\frac{\beta_1 - \beta_2}{2\pi \rad}\hat{j}_3 \big)  \compd   } \si^{(n)}(\compe,\compt) ,
\\
&\lam = \sum_{n}
\frac{1}{\sqrt{2 \pi \rad}} e^{\big( i \darn + (1 - \hat{j}_3)\frac{\beta_1}{2 \pi \rad} + \hat{j}_3 \frac{\beta_2}{ 2 \pi \rad} \big) \compd}
 \lam^{(n)} (\compe, \compt),
\\
&\olam = \sum_{n}\frac{1}{\sqrt{2 \pi \rad}} e^{\big( i \darn + (-1 - \hat{j}_3)\frac{\beta_1}{2 \pi \rad} + \hat{j}_3 \frac{\beta_2}{2 \pi \rad} \big) \compd}  \olam^{( n )} (\compe, \compt),
\end{align}
where $\hat{j}_3$ is the orbital angular momentum operator\footnote{When one generalizes it with the non-Abelian gauge group, one should replace $\pa_\compt$ by the covariant derivative defined by the precise flat connection $A_\fla^{(\pm)}$ in \eqref{j_3op} corresponding to the locus around which the fluctuation fields are expanded.}
\begin{align}
\hat{j}_3
=
-i \pa_\compt
.
\label{j_3op}
\end{align}
Then, one can get the sum of 2d Lagrangians $\Lcal^{{\rm 2d}\ (n)}$ of Kaluza-Klein fields labeled by $n$ after performing the integral along $\Sp^1$,
\begin{align}
\int d^3 x \sqrt{g_3}
\mathcal{L}
=
\int d^2 x \sqrt{g_2}
\sum_{n} \mathcal{L}^{2d \ (n)}.
\end{align}
The bosonic part and the fermionic part are as follows:
\begin{align}
\sqrt{g_2}
\mathcal{L}_{\text{boson}}^{{\rm 2d} \ (n)}
&=
\begin{pmatrix}
\A^{(-n)} \\
 A_3^{(-n)}\\
  \si^{(-n)}
\end{pmatrix}^{\rm T}
\! \! \! \! \!
\wedge 
*_2
\! 
\begin{pmatrix}
- (*_2d)^2 + \h^2 &- i \h d & - *_2d \frac{1}{f} \\
-i\h *_2d *_2 & - (*_2d)^2 & 0 \\
+\frac{1}{f} *_2 d  & 0 & - (*_2 d)^2 + \frac{1}{f^2} + \h^2 
\end{pmatrix}
\begin{pmatrix}
\A^{(n)} \\
 A_3^{(n)}\\
  \si^{(n)}
\end{pmatrix}
,
\label{vecbosd}
\\
\mathcal{L}_{\text{fermion}}^{{\rm 2d} \ (n)}
&=
\olam^{(-n)}
\Big(
i \g^i \D_i
+  \g_3 \big( \h + \frac{i}{2 l} \tri \big)
\Big)
\lam^{(n)}
,
\label{vecferd}
\end{align}
where $*_2$ is the Hodge star of $\Rp_b^2$, and the exterior derivative $d$ and the gauge field $\A^{(n)}$ are 1-forms on $\Rp_b^2$. The symbol $\h$ represents an operator defined by
\begin{align}
\h = -  \Big( \darn + \frac{i}{l} \tri \hat{j}_3 \Big).
\end{align}
The Lagrangians \eqref{vecbosd} and \eqref{vecferd} are quite similar to the ones  on $\Rp_b^2$ in \cite{Kim:2013ola} by identifying $\h \sim \alpha \cdot \sigma$. Although, in the fermionic term \eqref{vecferd}, a slightly different contribution exists, we can do the same procedure performed in the Appendix in \cite{Kim:2013ola}. See Appendix \ref{1-loop} for details.

\paragraph{One-loop determinant}
The final result is
\begin{align}
& \Zcal_\text{1-loop}^{\text{vector}(+)} (x) = \Zcal_\text{1-loop}^{\text{vector}(-)} (x)=
\mathcal{Z}_\text{1-loop}^{\text{vector}} (x)
=
x^{+\frac{1}{4}}
\exp{\Big(
\sum_{m=1}^\infty
\frac{1}{m}
f_{\text{vector}}(x^m)
\Big)},
\label{vec1loop}
\\
&f_{\text{vector}}(x)
=
\frac{x^2}{1-x^4}
- \frac{x^4}{1-x^4},
\label{vecind}
\end{align}
where the prefactor preceding the exponent is the \textit{Casimir energy} explained in detail in Appendixes \ref{1-loop} and \ref{Formula}. As the end of the discussion here, we would like to mention the origin of the U$(1)$ vector one-loop determinant. Intuitively, it is concluded as the difference of spins of bosonic and fermionic fields. More precisely, $Z_{2}$-holonomy splits the eigenvalues of $\hat{j}_{3}$ into two sets of integers which are assigned to each sector according to the spins of the fields. As a result, the eigenvalues run for odd integers in the bosonic sector and even integers in the fermionic sector under the B-parity condition \eqref{vecp}. This mismatch of the eigenvalues leads to the nontrivial one-loop determinant for U$(1)$ gauge group. The readers can see this explicitly also in Appendix \ref{Appvec}.

\subsection{Singlet matter multiplet contribution}
Second, we have to know the locus of the matter Lagrangian $\mathcal{L}_{\text{mat}}$ \eqref{mat}. However, it is somewhat trivial because the configuration is realized by turning off all fields in the matter multiplet. Therefore, by expanding around it, we get the following linearized Lagrangians:
\begin{align}
&\mathcal{L}_{\text{boson}}
=
g^{\mu \nu}
 \D_\mu^{\bold{A}} \ophi \D_\nu^{\bold{A}} \phi  
 - \frac{\Delta}{f}    \ophi \D_3^{\bold{A}} \phi  
- \frac{\Delta}{2f^2}  \ophi \phi 
+ \frac{\Delta}{4}R \ophi \phi
-\frac{\Delta-1}{f}v^i \ophi  \D_i^{\bold{A}} \phi
,
\label{bos2}
\\
&\mathcal{L}_{\text{fermion}}
=
-i  ( \opsi \g^\mu \D_\mu^{\bold{A}}\psi)
- \frac{i\Delta}{2f} ( \opsi \g_3 \psi)  
-i\frac{\Delta-1}{2f} v^i (\opsi \g_i \psi)
-i\frac{\Delta-1}{2f} \om (\opsi \psi)
.
\label{fer2}
\end{align}
Here, the superscript $\bold{A}$ means the covariant derivative \eqref{covderiv} defined with the locus value of the gauge field \eqref{locus}.

\paragraph{3d to 2d}
By expanding Kaluza-Klein modes first, we can get the 2d action as well as the case of the vector multiplet. In order to preserve SUSY, we have to read the precise boundary conditions from the fugacities in the index \eqref{indexdef}:
\begin{align}
&
\phi(\compe, \compt, \compd)
=
\sum_{n=-\infty}^\infty  \frac{1}{\sqrt{2 \pi}}e^{\big( i \darn+ (-\Delta-\hat{j}_3)\frac{\beta_1}{2 \pi \rad} + \hat{j}_3 \frac{\beta_2}{2 \pi \rad}-\frac{i \fl \mu}{2 \pi \rad}  \big) \compd} \phi^{(n)} (\compe, \compt) ,
\label{kkbcm1} \\
&
\ophi(\compe, \compt, \compd)
=
\sum_{n=-\infty}^\infty \frac{1}{\sqrt{2 \pi}} e^{\big( i \darn+ (+\Delta-\hat{j}_3)\frac{\beta_1}{2 \pi \rad} + \hat{j}_3 \frac{\beta_2}{2 \pi \rad}+\frac{i \fl \mu}{2 \pi \rad}  \big) \compd} \ophi^{(n)} (\compe, \compt) ,
\label{kkbcm2} \\
&
\psi(\compe, \compt, \compd)
=
\sum_{n=-\infty}^\infty \frac{1}{\sqrt{2 \pi}} e^{\big(i \darn+(-\Delta+1-\hat{j}_3)\frac{\beta_1}{2 \pi \rad} + \hat{j}_3 \frac{\beta_2}{2 \pi \rad}-\frac{i\fl \mu}{2 \pi \rad} \big) \compd} \psi^{(n)} (\compe, \compt) 
,
\label{kkbcm3} \\
&
\opsi(\compe, \compt, \compd)
=
\sum_{n=-\infty}^\infty \frac{1}{\sqrt{2 \pi}} e^{\big(i \darn+(+\Delta-1-\hat{j}_3)\frac{\beta_1}{2 \pi \rad} + \hat{j}_3 \frac{\beta_2}{2 \pi \rad}+\frac{i\fl \mu}{2 \pi \rad} \big) \compd} \opsi^{(n)} (\compe, \compt) 
,
\label{kkbcm4}
\end{align}
where $\hat{j}_3$ is the orbital angular momentum operator
\begin{align}
\hat{j}_3 = -i \Big( \pa_\compt - i \ch A^\fla_\compt \Big).
\end{align}
Note that there is a nontrivial contribution from the gauge field on the locus because the matter multiplet couples with the vector multiplet via the gauge symmetry. This effect is absent in the vector multiplet's case because it is neutral when the gauge group is U$(1)$.

Now, once we perform the integral over $\Sp^1$ as for the vector multiplet, we can get 2d Lagrangians,
\begin{align}
\mathcal{L}_{\text{boson}}^{{\rm 2d} \ (n)}
&=
\ophi^{(-n)}
\Big( - g^{ij}  \D_i^{A_\fla} \D_j^{A_\fla}
+(\p - i \frac{\Dl}{2 l}\tri )^2 
+ \frac{\Delta^2 - 2 \Delta}{4f^2}
+ \frac{\Delta}{4}R
-\frac{\Delta-1}{f}v^i   \D_i^{A_\fla}
 \Big) \phi^{(n)},
 \label{matbosd}
 \\
\mathcal{L}_{\text{fermion}}^{{\rm 2d} \ (n)}
&=
\opsi^{(-n)}
\Big(
-i  \g^i \D_i^{A_\fla}
- \g_3 (\p - i \frac{\Dl - 1}{2 l}\tri )
- i \frac{1}{2f} \g_3 
-i\frac{\Delta-1}{2f}v^i  \g_i 
-i\frac{\Delta-1}{2f} \om 
\Big)
\psi^{(n)},
 \label{matferd}
\end{align}
where the symbol $\p$ represents an operator defined by
\begin{align}
\p = 
-( n + \frac{i}{l}\tri \hat{j}_3)
+ \frac{ \ch \theta + \fl \mu }{2 \pi \rad}  
.
\end{align}
The Lagrangians \eqref{matbosd} and \eqref{matferd} are also similar to the ones on $\Rp_b^2$ in \cite{Kim:2013ola} by identifying $\p \sim \si$. As we can see in the fermionic part of the Lagrangian for the vector multiplet, there are also distinctions between \eqref{matbosd}, \eqref{matferd} and the corresponding ones in \cite{Kim:2013ola}. Even with these extra terms, we can perform exact calculations. See more details in Appendix \ref{1-loop}.

\paragraph{One-loop determinant}
The final result is
\begin{align}
&\mathcal{Z}_\text{1-loop}^{\Dl (+)} ( e^{i \ch \theta}, x, \alpha^\fl )
=
x^{+\frac{\Delta -1}{4}} e^{+ \frac{i}{4} \ch \theta} \alpha^{+\frac{1}{4} \fl}
\exp{\Big(
\sum_{m=1}^\infty
\frac{1}{m}
f^{(+)}_{\text{matter}}( 
e^{im {\ch} \theta} ,x^m, \alpha^{m \fl})
\Big)},
\label{mat+1loop}
\\
&
f^{(+)}_{\text{matter}}( e^{i {\ch} \theta} ,x, \alpha^{\fl})
=
e^{+i \ch \theta} \alpha^{+\fl} \frac{x^\Delta}{1-x^4}
- 
e^{-i \ch\theta} \alpha^{-\fl} \frac{x^{2-\Delta}}{1-x^4}
\label{mat+ind}
\end{align}
for the even holonomy sector which gives $e^{i \oint_\g (\ch A_\fla +  \fl B_\fla^{\text{flavor}})} =+1$. The other final form is
\begin{align}
&\mathcal{Z}_\text{1-loop}^{\Dl (-)} ( e^{i \ch \theta}, x, \alpha^\fl )
=
x^{-\frac{\Delta -1}{4}} e^{- \frac{i}{4} \ch\theta} \alpha^{-\frac{1}{4}\fl}
\exp{\Big(
\sum_{m=1}^\infty
\frac{1}{m}
f^{(-)}_{\text{matter}}(e^{im \ch \theta} ,x^m, \alpha^{m \fl})
\Big)},
\label{mat-1loop}
\\
&
f^{(-)}_{\text{matter}}(e^{i \ch\theta} ,x, \alpha^{\fl})
=
e^{+i \ch\theta} \alpha^{+\fl} \frac{x^{2+\Delta}}{1-x^4}
- 
e^{-i \ch\theta} \alpha^{-\fl} \frac{x^{4-\Delta}}{1-x^4}
\label{mat-ind}
\end{align}
when we have the odd holonomy sector $e^{i \oint_\g (\ch A_\fla +  \fl B_\fla^{\text{flavor}})} =-1$.

\subsection{Doublet matter multiplet contribution}
If we have a doublet matter multiplet constructed from two matter multiplets
\begin{align}
\Phi_1,
\quad
\Phi_2
\end{align}
with the matrix \eqref{XY},
\begin{align}
{\bf M} = {\bf N} =
\begin{pmatrix}
0 & 1\\
1 & 0
\end{pmatrix}
\end{align}
in the parity condition \eqref{ematp}, then, we can regard them as one matter multiplet on $\Sp_b^2 \times \Sp^1$ as commented above.

\paragraph{One-loop determinant}
Therefore, we can get the corresponding one-loop determinant just by quoting the one in the zero-monopole sector on $\Sp_b^2 \times \Sp^1$ \cite{Imamura:2011su}:
\begin{align}
&\mathcal{Z}_\text{1-loop}^{\Dl (2)} ( e^{i \ch \theta}, x, \alpha^\fl )
=
\exp{\Big(
\sum_{m=1}^\infty
\frac{1}{m}
f^{(2)}_{\text{matter}}(e^{im \ch \theta} ,x^m, \alpha^{m \fl})
\Big)},
\label{mat(2)1loop}
\\
&
f^{(2)}_{\text{matter}}(e^{i \ch\theta} ,x, \alpha^{\fl})
=
e^{+i \ch\theta} \alpha^{+\fl} \frac{x^{\Delta}}{1-x^2}
- 
e^{-i \ch\theta} \alpha^{-\fl} \frac{x^{2-\Delta}}{1-x^2}
\label{mat(2)ind}.
\end{align}
Note that there is $(1-x^2)$ in the denominator of \eqref{mat(2)ind} different from $(1-x^4)$ in \eqref{mat+ind} and \eqref{mat-ind}.

\subsection{Formulas of the SCI on $\Rp^2_b \times \Sp^1$}
Now, we summarize the relevant formulas of the SCI on $\Rp^2_b \times \Sp^1$ for later use in the check of 3d mirror symmetry. We specify with two types of theories. The first one is a class of matter theories. The second one is a class of U$(1)$ gauge theories. Before proceeding to the details of the formulas, we rewrite one-loop determinants in \eqref{vec1loop}, \eqref{mat+1loop}, and \eqref{mat-1loop} as more convenient forms. We now focus on the exponential part called the \textit{plethystic exponential} of the one-loop determinant of the vector multiplet \eqref{vec1loop}. It can be rewritten as follows. We use a geometric series for the one-particle index \eqref{vecind} and perform the sum over $m$. Then, the plethystic exponential becomes
\begin{align}
\exp \lp \sum_{m \geq 1} \frac{1}{m} f_{\rm vector} \lp x^{m} \rp \rp
&= \exp \lp \sum_{k \geq 0} \lc \log ( 1 - x^{4} x^{4 k} ) - \log ( 1 - x^{2} x^{4 k} ) \rc \rp \non \\ 
&= \prod_{k \geq 0} \frac{\lp 1 - x^{4} x^{4 k} \rp}{\lp 1 - x^{2} x^{4 k} \rp} \non \\ 
&= \frac{( x^{4} ; x^{4} )_{\infty}}{( x^{2} ; x^{4} )_{\infty}}, 
\end{align}
where we use the $q$-\textit{shifted factorial} defined by \cite{GR}
\begin{align} 
( Z ; q )_{n} := \lc
	\begin{aligned}
	& 1 && \mbox{for } n = 0, \\[.75em] 
	& \prod_{k = 0}^{n - 1} ( 1- Z q^{k} ) && \mbox{for } n \geq 1, \\ 
	& \prod_{k = 1}^{- n} ( 1- Z q^{- k} )^{- 1} && \mbox{for } n \leq - 1, 
	\end{aligned} \right.
\end{align}
where $Z$ and $q$ are complex numbers and $( Z ; q )_{\infty} := \lim_{n \to \infty} ( Z ; q )_{n}$ with $0 < | q | < 1$. For simplicity, we will use the notation
\begin{align}
( Z_{1}, Z_{2}, \cdots, Z_{r} ; q )_{\infty} := ( Z_{1} ; q )_{\infty} ( Z_{2} ; q )_{\infty} \cdots ( Z_{r}; q )_{\infty}.
\end{align}
The plethystic exponential of \eqref{mat+1loop}, \eqref{mat-1loop}, and \eqref{mat(2)1loop} can be written in the same manner with the $q$-shifted factorial,
\begin{align} 
\exp \lp \sum_{m \geq 1} \frac{1}{m} f_{\rm matter}^{(+)} \lp z^{\ch m}, x^{m}, \alpha^{\fl m} \rp \rp
&= \frac{( z^{- \ch} \alpha^{- \fl} x^{(2 - \Dl)} ; x^{4} )_{\infty}}{( z^{+ \ch} \alpha^{+ \fl} x^{\Dl} ; x^{4} )_{\infty}}, \\ 
\exp \lp \sum_{m \geq 1} \frac{1}{m} f_{\rm matter}^{(-)} \lp z^{\ch m}, x^{m}, \alpha^{\fl m} \rp \rp
&= \frac{( z^{- \ch} \alpha^{- \fl} x^{(4 - \Dl)} ; x^{4} )_{\infty}}{( z^{+ \ch} \alpha^{+ \fl} x^{( 2 + \Dl )} ; x^{4} )_{\infty}}, \\ 
\exp \lp \sum_{m \geq 1} \frac{1}{m} f_{\rm matter}^{(2)} \lp z^{\ch m}, x^{m}, \alpha^{\fl m} \rp \rp
&= \frac{( z^{- \ch} \alpha^{- \fl} x^{(2 - \Dl)} ; x^{2} )_{\infty}}{( z^{+ \ch} \alpha^{+ \fl} x^{\Dl} ; x^{2} )_{\infty}},  
\end{align}
where we define $z := e^{i \theta}$. Combining the Casimir energy \eqref{casivv}, \eqref{casip}, and \eqref{casim} together, we have the following one-loop determinants for each multiplet:
\begin{align} 
	\begin{aligned}
	\Zcal_\text{1-loop}^{\rm vector} (x) &= x^{+ \frac{1}{4}} \frac{( x^{4} ; x^{4} )_{\infty}}{( x^{2} ; x^{4} )_{\infty}}, \\ 
	\Zcal_\text{1-loop}^{\Dl (+)} ( z^{\ch}, x, \alpha^\fl )
	&= x^{+ \frac{\Dl - 1}{4}} z^{+ \frac{1}{4} \ch} \alpha^{+ \frac{1}{4} \fl}
	\frac{( z^{- \ch} \alpha^{- \fl} x^{(2 - \Dl)} ; x^{4} )_{\infty}}{( z^{+ \ch} \alpha^{+ \fl} x^{\Dl} ; x^{4} )_{\infty}}, \\ 
	\Zcal_\text{1-loop}^{\Dl (-)} ( z^{\ch}, x, \alpha^\fl )
	 &= x^{- \frac{\Dl - 1}{4}} z^{- \frac{1}{4} \ch} \alpha^{- \frac{1}{4} \fl}
	\frac{( z^{- \ch} \alpha^{- \fl} x^{(4 - \Dl)} ; x^{4} )_{\infty}}{( z^{+ \ch} \alpha^{+ \fl} x^{( 2 + \Dl )} ; x^{4} )_{\infty}}, \\ 
	\Zcal_\text{1-loop}^{\Dl (2)} ( z^{\ch}, x, \alpha^\fl )
	 &= 
	\frac{( z^{- \ch} \alpha^{- \fl} x^{(2 - \Dl)} ; x^{2} )_{\infty}}{( z^{+ \ch} \alpha^{+ \fl} x^{\Dl} ; x^{2} )_{\infty}}. 
	\end{aligned}
	\label{1-loopindex}
\end{align}

\paragraph{SCI for matter theory}
First of all, for later use, we consider a theory constructed only by multimatter multiplets as follows:
\begin{align}
&\bullet \text{Singlet matter multiplets with the upper sign in \eqref{sematp}} \hspace{-1.25em} &&: \ \Phi^a_+, \overline{\Phi}^a_+,
&&a=1,2,\dots, N_f^{(+)}.
\notag 
\\
&\bullet \text{Singlet matter multiplets  with the lower sign in \eqref{sematp}} \hspace{-1.25em} &&: \ \Phi^{b}_-, \overline{\Phi}^{b}_-,
&&b=1,2,\dots, N_f^{(-)}.
\notag 
\\
&\bullet \text{Doublet matter multiplets with the parity by \eqref{XY}} \hspace{-1.25em} &&: \ \Phi_{1,2}^A, \overline{\Phi}_{1,2}^A,
&&A=1,2,\dots, N_f^{(2)}.
\notag 
\end{align}
We can turn on arbitrary superpotentials. However, we assume here that we turn a certain superpotential which restricts the flavor symmetries to one global U$(1)$ symmetry, and we denote the corresponding fugacity by $\alpha$. In this case, the SCI does not contain any summation or integral over the configuration of the gauge field, and our result is 
\begin{align}
\mathcal{I}_{} (x,{\alpha})
=
\prod_{a=1}^{N_f^{(+)}}
\mathcal{Z}_{\text{1-loop}} ^{\Dl_a 
 (+)} (1, x, {\alpha}^{\fl_a})
\prod_{b=1}^{N_f^{(-)}}
\mathcal{Z}_{\text{1-loop}} ^{\Dl_{b} 
(-)} (1, x, {\alpha}^{\fl_{b}})
\prod_{A=1}^{N_f^{(2)}}
\mathcal{Z}_{\text{1-loop}} ^{\Dl_A 
(2)} (1, x, {\alpha}^{\fl_A}),
\label{nogSCIr}
\end{align}
where flavor charges $\fl_a, \fl_b, \fl_A$ for the global U$(1)$ symmetry and R-charges $\Dl_a, \Dl_b, \Dl_A$ are assigned to the matter multiplets $\Phi_+^a$, $\Phi_-^b$, and $\Phi_{1,2}^A$, respectively.

\paragraph{SCI for gauge theory}
The second example is a U$(1)$ gauge theory constructed from
\begin{align}
&\bullet \text{a vector multiplet} \hspace{-1em} &&: \ V,
\notag
\\
&\bullet \text{singlet matter multiplets with the upper sign in \eqref{sematp}} \hspace{-1em} &&: \ \Phi^a_+, \overline{\Phi}^a_+,
&&a=1,2,\dots, N_f^{(+)},
\notag 
\\
&\bullet \text{singlet matter multiplets  with the lower sign in \eqref{sematp}} \hspace{-1em} &&: \ \Phi^{b}_-, \overline{\Phi}^{b}_-,
&&b=1,2,\dots, N_f^{(-)},
\notag 
\\
&\bullet \text{doublet matter multiplets with the parity by \eqref{XY}} \hspace{-1em} &&: \ \Phi_{1,2}^A, \overline{\Phi}_{1,2}^A,
&&A=1,2,\dots, N_f^{(2)}.
\notag
\end{align}
In addition to the global U$(1)$ symmetry and R-charges, we assign U$(1)$ gauge charges $\ch_a , \ch_{b}, \ch_A \in \mathbb{Z}$ for  $\Phi_+^a$, $\Phi_-^b, \Phi_{1,2}^A$, respectively. As noted in \eqref{gaugecont}, we have a summation over two terms coming from two distinct loci $A_\fla = A_\fla^{(\pm)}$ and an integral over loci $\theta \in [0 , 2 \pi]$. With the results \eqref{vec1loop}, \eqref{mat+1loop}, and \eqref{mat-1loop} obtained by the localization, the one-loop determinants for the multiplets are given as follows:
\begin{align}
	\begin{aligned}
	V &\to
	\mathcal{Z}_{\text{1-loop}}^{\text{vector}} \text{ always in \eqref{1-loopindex}}, \\ 
	\Phi_+^a, \overline{\Phi}_+^a &\to
	\mathcal{Z}_{\text{1-loop}}^{\Dl_{a}} \text{ depending on the sign } e^{i \oint_\g (\ch_a A_\fla + \fl_a B_\fla^{a})} = + e^{i \oint_\g (\ch_a A_\fla )}, \non \\ 
	\Phi_-^b, \overline{\Phi}_-^b &\to
	\mathcal{Z}_{\text{1-loop}}^{\Dl_{b}} \text{ depending on the sign } e^{i \oint_\g (\ch_b A_\fla + \fl_b B_\fla^{b})} = - e^{i \oint_\g (\ch_b A_\fla )}, \non \\ 
	\Phi_{1,2}^A, \overline{\Phi}_{1,2}^A &\to
	\mathcal{Z}_{\text{1-loop}}^{\Dl_A (2)} \text{ always in \eqref{1-loopindex}}. \non 
	\end{aligned}
\end{align}
The one-loop determinant $\mathcal{Z}_{\text{1-loop}}^{\Dl (\pm)}$ for a singlet matter multiplet is slightly complicated, and let us give a detailed explanation here. For the $A_\fla=A_\fla^{(+)}$ sector, we have $e^{i \oint_\g (\ch A_\fla^+ )} = (e^{i \oint_\g A_\fla^+ } )^{\ch}  = (+1)^\ch =+1 $, and then we get the one-loop determinants for $\Phi_\pm$ as $\mathcal{Z}_{\text{1-loop}}^{\Delta (\pm)}$ in \eqref{1-loopindex}, respectively.

On the other hand, for the $A_\fla=A_\fla^{(-)}$ sector, we have $e^{i \oint_\g (\ch A_\fla^- )} = (e^{i \oint_\g A_\fla^- } )^{\ch}  = (-1)^\ch $, and its value depends on the parity of $\ch \in \mathbb{Z}$. For simplicity, we assume the following circumstances:
\begin{align}
\text{Each charge } \ch_a, \ch_b \text{ takes its value in odd integers}.
\label{condA}
\end{align}
This is satisfied in the latter part of this paper. Then, we get the one-loop determinants for $\Phi_\pm$ as $\mathcal{Z}_{\text{1-loop}}^{\Delta (\mp)}$ in \eqref{1-loopindex} with $A_\fla = A_\fla^{(-)}$ sector. Under the condition \eqref{condA}, we arrive at the following formula for the SCI by defining $z=e^{i \theta}$:
\begin{align}
&\mathcal{I}_{} (x,{\alpha})
\notag \\
&=
\frac{\mathcal{Z}_{\text{1-loop}} ^{\text{vector} } (x)}{(Sym)}
\oint_{C_0} \frac{d z}{2 \pi i z}
\lc
\prod_{a=1}^{N_f^{(+)}}
\mathcal{Z}_{\text{1-loop}} ^{\Dl_a 
(+)} (z^{ \ch_a}, x, {\alpha}^{\fl_a})
\prod_{b=1}^{N_f^{(-)}}
\mathcal{Z}_{\text{1-loop}} ^{\Dl_{b} 
(-)} (z^{\ch_{b} }, x, {\alpha}^{\fl_{b}})
\prod_{A=1}^{N_f^{(2)}}
\mathcal{Z}_{\text{1-loop}} ^{\Dl_A 
(2)} (z^{ \ch_A }, x, {\alpha}^{\fl_A}) \right.
\notag \\ & \left. \qquad \qquad \qquad \qquad \qquad \quad
+
\prod_{a=1}^{N_f^{(+)}}
\mathcal{Z}_{\text{1-loop}} ^{\Dl_a 
(-)} (z^{\ch_a}, x, {\alpha}^{\fl_a})
\prod_{b=1}^{N_f^{(-)}}
\mathcal{Z}_{\text{1-loop}} ^{\Dl_{b} 
(+)} (z^{\ch_{b}}, x, {\alpha}^{\fl_{b}})
\prod_{A=1}^{N_f^{(2)}}
\mathcal{Z}_{\text{1-loop}} ^{\Dl_A 
(2)} (z^{ \ch_A }, x, {\alpha}^{\fl_A})
\rc.
\label{gSCIr}
\end{align}
The integration contour $C_0$ is defined by $|z| = 1$ because $|z|=|e^{i \theta}| =1$ by definition. The symbol $(Sym)$ represents degrees of a redundant symmetry between two sectors $e^{i \oint_\g  A_\fla } = \pm 1$. If the first integrand is identical to the second one, it is 2. If not, it is 1.

\section{Abelian Mirror Symmetry} \label{AMS}
We start with the review of Abelian mirror symmetry for 3d $\Ncal = 2$ theories \cite{Intriligator:1996ex, Aharony:1997bx, Kapustin:1999ha, Tong:2000ky} with a single flavor. Then, we explain how this duality can be realized in terms of the SCIs for theories on $\mathbb{R}\mathbb{P}_{b}^{2} \times \mathbb{S}^{1}$ in the physical sense and provide the mathematically exact verification to it.

\subsection{Review of 3d mirror symmetry}
$\Ncal = 2$ mirror symmetry states the duality between the SQED and the XYZ model. From the renormalization group point of view, theses two theories are defined in the UV region and flow to the same IR fixed point. The $\Ncal = 2$ SQED has one vector multiplet $V$ and one flavor consisting of two chiral fields $Q$, $\tilde{Q}$ with charges $\ch = + 1, - 1$ under the U$(1)$ gauge group, respectively. This theory possesses extra U$(1)$ global symmetries: one is a topological U$(1)_{J}$, and the other is a flavor symmetry U$(1)_{A}$ with a charge $\fl = + 1$ which rotates $Q$ and $\tilde{Q}$ by the phase with the same weight as seen in Table \ref{cSQED}. On the other hand, the XYZ model is the theory containing three chiral fields\footnote{In the literature \cite{Imamura:2011su, Krattenthaler:2011da, Kapustin:2011jm}, they are named $q, \tilde{q}$, and $S$, respectively.} $X, Y$, and $Z$ interacting through the superpotential $\Wcal = XYZ$. This theory has two U$(1)$ global symmetries, named U$(1)_{V}$ and U$(1)_{A}$ in \cite{Kapustin:2011jm}, whose charges assigned on each field are shown in Table \ref{cXYZ}.

U$(1)_{J}$ and U$(1)_{A}$ in the SQED are identified with U$(1)_{V}$ and U$(1)_{A}$ in the XYZ model, respectively, and the currents $J_{A}$ associated with each U$(1)_{A}$ are mapped with flipping the sign (see Table \ref{mmap}). Furthermore, there exists the correspondence between the moduli spaces of those theories (at least on the flat space). The moduli parameters of the SQED are $Q \tilde{Q}$ characterizing the Higgs branch and $( \sg + i \rho )$ where $\rho$ is the dual photon defined by
\begin{align} 
\frac{1}{2} \e_{\mu \nu \rho} F^{\nu \rho} = \partial_{\mu} \rho.
\label{dualphoton}
\end{align}
The expectation values of two chiral superfields $e^{\lp \sg + i \rho \rp / e^{2}}, e^{- \lp \sg + i \rho \rp / e^{2}}$ ($e$ is a coupling constant) parametrize the corresponding regions of the Coulomb branch. In the context of mirror symmetry, we can identify $e^{\lp \sg + i \rho \rp / e^{2}}, e^{- \lp \sg + i \rho \rp / e^{2}},$ and $Q \tilde{Q}$ with $X, Y,$ and $Z$ on the moduli space of the XYZ model, respectively (Table \ref{mmap}). 

\begin{table}[t]
\vspace{-2em}
\begin{center}
\begin{tabular}{c}
	\begin{minipage}[t]{.5\hsize} 
		\caption{Charges in the SQED\label{cSQED}}
		\vspace{-.5em}
		\begin{center}
			\begin{tabular}{|c|c|c|c|c|} \hline
			& U$(1)$ & U$(1)_{J}$ & U$(1)_{A}$ & $\hat{R}$ \\ \hline \hline
			$Q$ & $+ 1$ & $0$ & $+ 1$ & $- \Dl$ \\ \hline
			$\tilde{Q}$ & $- 1$ & $0$ & $+ 1$ & $- \Dl$ \\ \hline
			\end{tabular}
		\end{center}
	\end{minipage}

	\begin{minipage}[t]{.5\hsize} 
		\caption{Charges in the XYZ model\label{cXYZ}}
		\vspace{-.5em}
		\begin{center}
			\begin{tabular}{|c|c|c|c|} \hline
			& U$(1)_{V}$ & U$(1)_{A}$ &$\hat{R}$ \\ \hline \hline
			$X$ & $+ 1$ & $+ 1$ & $- ( 1 - \Dl )$ \\ \hline
			$Y$ & $- 1$ & $+ 1$ & $- ( 1 - \Dl )$ \\ \hline
			$Z$ & 0 & $- 2$ & $- 2 \Dl$ \\ \hline
			\end{tabular}
		\end{center}
	\end{minipage}
\end{tabular}
\end{center}
\end{table}

\begin{table}[t] 
\caption{The mirror map\label{mmap}}
\vspace{-.5em}
\begin{center}
	\begin{tabular}{ccc} \hline
	SQED & & XYZ \\ \hline
	U$(1)_{J}$ & $\leftrightarrow$ & U$(1)_{V}$ \\
	U$(1)_{A}$, $J_{A}$ & $\leftrightarrow$ & U$(1)_{A}$, $- J_{A}$ \\
	$e^{\lp \sg + i \rho \rp / e^{2}}$, $e^{- \lp \sg + i \rho \rp / e^{2}}$ & $\leftrightarrow$ & $X$, $Y$ \\
	$Q \tilde{Q}$ & $\leftrightarrow$ & $Z$ \\ \hline
	\end{tabular}
\end{center}
\vspace{-1em}
\end{table}

We can also construct the $\Ncal = 4$ version of mirror symmetry. In the SQED, we introduce an adjoint (uncharged) chiral field $\tilde{S}$ coupling to $Q \tilde{Q}$. Similarly for the XYZ model, $\tilde{Z}$ is added via the superpotential $Z \tilde{Z}$ making $Z$ and $\tilde{Z}$ massive. We can obtain the (twisted) free theory with chiral fields $X$ and $Y$ by integrating out $Z$ and $\tilde{Z}$. The duality between those theories is referred to as $\Ncal = 4$ mirror symmetry.

Let us now consider gauging a flavor symmetry and denote a corresponding background gauge field by $B^{\rm flavor}$. In addition to $J_{A}$, there is a topological current\footnote{For non-Abelian theories, a topological current should be in the form $J_{T} = \ast \Tr F$.} $J_{T} = \ast F$ associated with U$(1)_{J}$ where $\ast$ is the Hodge star defined by a 3d metric. The flavor symmetry can be gauged by coupling $B^{\rm flavor}$ with $J_{T}$, which is the same thing as adding a BF term to the original action \cite{Brooks:1994nn, Kapustin:1999ha}. This fact can be employed to demonstrate mirror symmetry with general $N_{f}$ in terms of generalized indices \cite{Kapustin:2011jm}.

\subsection{Physical derivation on $\mathbb{R}\mathbb{P}_{b}^{2} \times \mathbb{S}^{1}$} \label{PhysicalDeriv}
In this subsection, we construct Abelian mirror symmetry on $\mathbb{R}\mathbb{P}_{b}^{2} \times \mathbb{S}^{1}$ from physical viewpoints. Here, we should note that U$(1)_{J}$ in the SQED and U$(1)_{V}$ in the XYZ model on $\mathbb{R}\mathbb{P}_{b}^{2} \times \mathbb{S}^{1}$ cannot be turned on. This is because, for U$(1)_{J}$, a BF term is parity odd under the B-parity condition as well as the Chern-Simons term (see \eqref{BF} and \eqref{csbfparity}). On the dual side, since $\sg$ receives the change of sign by the antipodal identification \eqref{vecp}, $X$ and $Y$ seem to be interchanged from each other from the mirror map (the third line of Table \ref{mmap}). However, this violates U$(1)_{V}$ because $X$ has its charge opposite to that of $Y$. This is why there do not exist variables in the SCIs parametrizing U$(1)_{J}$ and U$(1)_{V}$ in the latter argument.

\paragraph{SQED} 
The SCI should be the sum of the even and odd holonomy sector of the dynamical gauge field as described in \eqref{gSCIr}. We consider the following situations:
\begin{align}
& \bullet \text{the vector multiplet }V \text{ with the parity condition \eqref{vecp}, and}
\label{Vcond}
\\
& \bullet \text{the matter multiplets }Q, \tilde{Q} \text{ with the upper sign of the parity condition in \eqref{sematp}.}
\label{QtQcond}
\end{align}
Then, by taking into account the charge assignments in Table \ref{cSQED}, we get the SCI for the SQED immediately from the general formula \eqref{gSCIr} as
\begin{align}
& \Ical_{\rm SQED} ( x, \alpha )
\notag \\
&=
\Zcal_\text{1-loop}^{\text{vector}} (x) 
\oint_{C_0} \frac{d z}{2 \pi i z}
\lc
\Zcal_\text{1-loop}^{\Dl (+)} (z^{+1} x, \alpha)
\Zcal_\text{1-loop}^{\Dl (+)} (z^{-1}, x, \alpha)
+
\Zcal_\text{1-loop}^{\Dl (-)} (z^{+1} x, \alpha)
\Zcal_\text{1-loop}^{\Dl (-)} (z^{-1}, x, \alpha)
\rc \non \\[.5em] 
\label{01}
\end{align}
To make the SCI easy to deal with, we introduce new variables,
\begin{align}
q = x^{2}, \hspace{2em}
a = \alpha^{- 2 } x^{2 ( 1 - \Dl )}.
\end{align}
Then, we get the following representation:
\begin{align}
&\mathcal{I}_{\text{SQED}}(x,\alpha)
\notag \\
&= q^{+ \frac{1}{8}} \frac{( q^{2} ; q^{2} )_{\infty}}{( q ; q^{2} )_{\infty}}
\oint_{C_0} \frac{d z}{2 \pi i z} \lc a^{- \frac{1}{4}}
\frac{( z^{- 1} a^{+ \frac{1}{2}} q^{\frac{1}{2}}, z^{+ 1} a^{+ \frac{1}{2}} q^{\frac{1}{2}}; q^{2} )_{\infty}}
{( z^{+ 1} a^{- \frac{1}{2}} q^{\frac{1}{2}}, z^{- 1} a^{- \frac{1}{2}} q^{\frac{1}{2}} ; q^{2} )_{\infty}}
+  a^{+ \frac{1}{4}}
\frac{( z^{- 1} a^{+ \frac{1}{2}} q^{\frac{3}{2}}, z^{+ 1} a^{+ \frac{1}{2}} q^{\frac{3}{2}} ; q^{2} )_{\infty}}
{( z^{+ 1} a^{- \frac{1}{2}} q^{\frac{3}{2}}, z^{- 1} a^{- \frac{1}{2}} q^{\frac{3}{2}} ; q^{2} )_{\infty}} \rc, 
\label{SQED1}
\end{align}
We follow the way of \cite{Krattenthaler:2011da, Kapustin:2011jm} to perform the above integrals. We start to handle the first integral in \eqref{SQED1}. There are many single poles coming from the origin and the $q$-shifted factorial. Those poles can be separated into the set inside and outside the unit circle $C_0$. We set $|q| < 1$ for the convergence of the $q$-shifted factorial and assume $| a^{- \frac{1}{2}} q^{\frac{1}{2}} | < 1$. Then the poles we should take into account are the ones inside the unit circle,
\begin{align} 
z = a^{- \frac{1}{2}} q^{\frac{1}{2} + 2 j}, \hspace{2em}
j = 0, 1, 2, \cdots.
\label{poles}
\end{align}
We can relax the assumption by analytic continuation after obtaining the final result. In addition, we assume $0 < \Dl < 1$ because picking up the pole on the origin leads to the infinite product
\begin{align}
\prod_{j = 0}^{\infty} a = \prod_{j = 0}^{\infty} \alpha^{- 2} q^{( 1 - \Dl )}
\label{origin}
\end{align}
except the Casimir energy. $\alpha$ is just a phase, and this product converges to zero if and only if we impose such condition on $\Dl$ with $|q| < 1$. Eventually, we can ignore the contribution of the pole on the origin. Then, the integral over $z$ with these assumptions gives the sum over residues from \eqref{poles} as
\begin{align}
a^{- \frac{1}{4}} \sum_{j \geq 0}
\frac{( a q^{- 2 j}, q^{1 + 2 j} ; q^{2} )_{\infty}}
{( a^{- 1} q^{1 + 2 j}, q^{2} ; q^{2} )_{\infty}}
\frac{1}{( q^{- 2 j} ; q^{2} )_{j}}.
\label{residue2}
\end{align}
We also rewrite the sum over $j$ as follows. The dummy index $j$ in arguments of the $q$-shifted factorial can be subtracted outside such as
\begin{align}
( a q^{- 2 j} ; q^{2} )_{\infty}
&= ( - 1 )^{j} a^{j} q^{- j ( j + 1 )} ( a^{- 1} q^{2} ; q^{2} )_{j} ( a ; q^{2} )_{\infty}, \\ 
( q^{1 + 2 j} ; q^{2} )_{\infty}
&= \frac{( q ; q^{2} )_{\infty}}{( q ; q^{2} )_{j}}. 
\end{align}
With above expressions, \eqref{residue2} reduces to
\begin{align}
& a^{- \frac{1}{4}} \frac{( a, q ; q^{2} )_{\infty}}{( a^{- 1} q, q^{2} ; q^{2} )_{\infty}} \sum_{j \geq 0}
\frac{( a^{- 1} q^{2}, a^{- 1} q ; q^{2} )_{j}}{( q ; q^{2} )_{j}} \frac{a^{j}}{(q^{2} ; q^{2} )_{j}} \non \\ 
=& \hspace{.25em} a^{- \frac{1}{4}} \frac{( a, q ; q^{2} )_{\infty}}{( a^{- 1} q, q^{2} ; q^{2} )_{\infty}}
\ {}_{2}\varphi_{1} \lp a^{- 1} q^{2}, a^{- 1} q ; q ; q^{2}, a \rp, 
\end{align}
where we use the \textit{basic hypergeometric series} defined by \cite{GR}
\begin{align} 
{}_{r}\varphi_{s} \lp \alpha_{1}, \alpha_{2}, \cdots, \alpha_{r} ; \beta_{1}, \cdots, \beta_{s} ; q, z \rp
= \sum_{j \geq 0} \frac{( \alpha_{1}, \alpha_{2}, \cdots, \alpha_{r}; q )_{j}}{( \beta_{1}, \cdots, \beta_{s}; q )_{j}} \frac{z^{j}}{( q; q)_{j}} \lc ( - 1 )^{j} q^{\frac{1}{2} j ( j - 1 )} \rc^{1 + s - r}.
\end{align}
The convergence radius of the basic hypergeometric series is $\infty$, $1$, or $0$ for $r - s < 1$, $r - s = 1$, or $r - s > 1$, respectively. Now, we proceed the same way for the second integral in \eqref{SQED1}. We pick up the poles again inside the unit circle $C_{0}$,
\begin{align} 
z = a^{- \frac{1}{2}} q^{\frac{3}{2} + 2 j}, \hspace{2em}
j = 0, 1, 2, \cdots,
\label{poles2}
\end{align}
and then the sum over residues in terms of the basic hypergeometric series becomes
\begin{align}
a^{+ \frac{1}{4}}
\frac{( a, q^{3} ; q^{2} )_{\infty}}{( a^{- 1} q^{3}, q^{2} ; q^{2} )_{\infty}}
\ {}_{2}\varphi_{1} \lp a^{- 1} q^{2}, a^{- 1} q^{3} ; q^{3} ; q^{2}, a \rp,
\end{align}
where the residue of the origin is not included as discussed in \eqref{origin}. Thus, \eqref{SQED1} results in
\begin{align}
\Ical_{\rm SQED} ( x, \alpha ) 
&= q^{+ \frac{1}{8}} \frac{( q^{2} ; q^{2} )_{\infty}}{( q ; q^{2} )_{\infty}}
\lc a^{- \frac{1}{4}}
\frac{( a, q ; q^{2} )_{\infty}}{( a^{- 1} q, q^{2} ; q^{2} )_{\infty}}
\ {}_{2}\varphi_{1} \lp a^{- 1} q^{2}, a^{- 1} q ; q ; q^{2}, a \rp \right. \non \\ 
&\hspace{6.75em} + \left. a^{+ \frac{1}{4}}
\frac{( a, q^{3} ; q^{2} )_{\infty}}{( a^{- 1} q^{3}, q^{2} ; q^{2} )_{\infty}}
\ {}_{2}\varphi_{1} \lp a^{- 1} q^{2}, a^{- 1} q^{3} ; q^{3} ; q^{2}, a \rp \rc. 
\label{amirror}
\end{align}
In terms of original variables, the index \eqref{amirror} is given by
\begin{align}
\Ical_{\rm SQED} ( x, \alpha ) 
&= x^{+ \frac{1}{4}} \frac{( x^{4} ; x^{4} )_{\infty}}{( x^{2} ; x^{4} )_{\infty}} \times \non \\ 
&\hspace{1em} \lc x^{+ \frac{\Dl - 1}{2}} \alpha^{+ \frac{1}{2}}
\frac{( \alpha^{- 2} x^{2 ( 1 - \Dl )}, x^{2} ; x^{4} )_{\infty}}{(\alpha^{+ 2} x^{2 \Dl}, x^{4} ; x^{4} )_{\infty}}
\ {}_{2}\varphi_{1} \lp \alpha^{+ 2} x^{2 ( \Dl + 1 )}, \alpha^{+ 2} x^{2 \Dl} ; x^{2} ; x^{4}, \alpha^{- 2} x^{2 ( 1 - \Dl )} \rp \right. \non \\ 
&\hspace{.7em} + \left. x^{- \frac{\Dl - 1}{2}} \alpha^{- \frac{1}{2}}
\frac{( \alpha^{- 2} x^{2 ( 1 - \Dl )}, x^{6} ; x^{4} )_{\infty}}{(\alpha^{+ 2} x^{2 ( 2 + \Dl )}, x^{4} ; x^{4} )_{\infty}}
\ {}_{2}\varphi_{1} \lp \alpha^{+ 2} x^{2 ( \Dl + 1 )}, \alpha^{+ 2} x^{2 ( 2 + \Dl )} ; x^{6} ; x^{4}, \alpha^{- 2} x^{2 ( 1 - \Dl )} \rp \rc. 
\label{amirror2}
\end{align}

\paragraph{XYZ model} 
We must determine the suitable pair of B-parity condition for three chiral fields which is ``mirror" to the pair of parity condition \eqref{Vcond} and \eqref{QtQcond} for the SQED to obtain the correct results. As described above, $X$ turns into $Y$ under the antipodal identification, and vice versa. We assume that this observation also holds for the quantum fluctuations of the XYZ model. Then, we set the B-parity condition for these fields as
\begin{align}
	\begin{aligned}
	X ( \pi - \vartheta, \pi + \varphi, y ) &= Y ( \vartheta, \varphi, y ), \\ 
	Y ( \pi - \vartheta, \pi + \varphi, y ) &= X ( \vartheta, \varphi, y ), \\ 
	Z ( \pi - \vartheta, \pi + \varphi, y ) &= Z ( \vartheta, \varphi, y ). 
	\end{aligned}
\end{align}
This means that $X$ and $Y$ form the doublet that appears in the previous section and provides the contribution of a single field on $\mathbb{S}_{b}^{2} \times \mathbb{S}^{1}$. With the charge assignments summarized in Table \ref{cXYZ}, we get the following contribution from $X$ and $Y$:
\begin{align} 
\Zcal_\text{1-loop}^{
	(1-\Dl) (2)}
	( 1 , x , {\tilde{\alpha}}^{\fl} )
=
\frac{( \tilde{\alpha}^{- 1} x^{( 1 + \Dl )} ; x^{2} )_{\infty}}{( \tilde{\alpha}^{+ 1} x^{( 1 - \Dl )} ; x^{2} )_{\infty}}
= \frac{( \tilde{a}^{- \frac{1}{2}} q ; q )_{\infty}}{( \tilde{a}^{+ \frac{1}{2}} ; q )_{\infty}},
\end{align}
where we define a fugacity $\tilde{\alpha}$ for the U$(1)_A$ global symmetry in the XYZ model and also $\tilde{a} := \tilde{\alpha}^{+ 2} x^{2 ( 1 - \Dl )}$ for later use. On the other hand, because $Z$ is a scalar invariant under the antipodal identification, the contribution of $Z$ corresponds to that of the even holonomy sector in the matter multiplet with the R-charge $- 2 \Dl$,
\begin{align} 
\Zcal_\text{1-loop}^{
	(2\Dl) (+)}
	( 1 , x , {\tilde{\alpha}}^\fl )
=
x^{+ \frac{2 \Dl - 1}{4}} \tilde{\alpha}^{- \frac{1}{2}} \frac{( \tilde{\alpha}^{+ 2} x^{2 ( 1 - \Dl )} ; x^{4} )_{\infty}}{( \tilde{\alpha}^{- 2} x^{2 \Dl} ; x^{4} )_{\infty}}
=
q^{+ \frac{1}{8}} \tilde{a}^{- \frac{1}{4}} \frac{( \tilde{a} ; q^{2} )_{\infty}}{( \tilde{a}^{- 1} q ; q^{2} )_{\infty}}.
\label{ss}
\end{align}
Because of the formula in \eqref{nogSCIr},
the SCI for the XYZ model results in
\begin{align} 
\Ical_{\rm XYZ} ( x, \tilde{\alpha} )
= q^{+ \frac{1}{8}} \tilde{a}^{- \frac{1}{4}}
\frac{( \tilde{a}^{- \frac{1}{2}} q ; q )_{\infty}}{( \tilde{a}^{+ \frac{1}{2}} ; q )_{\infty}}
\frac{( \tilde{a} ; q^{2} )_{\infty}}{( \tilde{a}^{- 1} q ; q^{2} )_{\infty}}.
\label{xyz}
\end{align}
Equivalently, \eqref{xyz} with original variables is written by
\begin{align} 
\Ical_{\rm XYZ} ( x, \tilde{\alpha} )
= x^{+ \frac{2 \Dl - 1}{4}} \tilde{\alpha}^{- \frac{1}{2}}
\frac{( \tilde{\alpha}^{- 1} x^{( 1 + \Dl )} ; x^{2} )_{\infty}}{( \tilde{\alpha}^{+ 1} x^{( 1 - \Dl )} ; x^{2} )_{\infty}}
\frac{( \tilde{\alpha}^{+ 2} x^{2 ( 1 - \Dl )} ; x^{4} )_{\infty}}{( \tilde{\alpha}^{- 2} x^{2 \Dl} ; x^{4} )_{\infty}}.
\label{xyz2}
\end{align}

In the expressions of the SCIs, the usual mirror map for a flavor symmetry is realized by the identification $\alpha \sim \tilde{\alpha}^{- 1}$, or, equivalently, $a \sim \tilde{a}$ in our notation. Accordingly, we declare $\Ncal = 2$ Abelian mirror symmetry on $\mathbb{RP}_{b}^{2} \times \mathbb{S}^{1}$ as the equality
\begin{align} 
\Ical_{\rm SQED} ( x, \alpha ) = \Ical_{\rm XYZ} ( x, \tilde{\alpha}^{- 1} ).
\label{amirror3}
\end{align}
Note that \eqref{amirror3} should be true with an arbitrary $\Dl$, whereas the R-charge in theories without anomalous dimensions must take the canonical value as mentioned in \cite{Imamura:2011su}. In the next subsection, we will show the mathematically rigorous proof of \eqref{amirror3}.

\paragraph{$\Ncal = 4$ mirror symmetry} 
As explained above, we can obtain $\Ncal = 4$ mirror symmetry by introducing an adjoint chiral field $\tilde{Z}$. In the XYZ model, the fact that the superpotential $Z \tilde{Z}$ must be uncharged for a flavor symmetry and have the R-charge $2$ determines the U$(1)_{A}$ charge and the R-charge of $\tilde{Z}$ to be $+ 2$ and $2 ( 1 + \Dl )$, respectively.
For the SCIs, the effect of $\tilde{Z}$ is identical with moving the contribution of $Z$ \eqref{ss} in the rhs of \eqref{amirror3} to the lhs. Concretely, we have the equality for $\Ncal = 4$ mirror symmetry as
\begin{align}
\ {}_{2}\varphi_{1} \lp a^{- 1} q^{2}, a^{- 1} q ; q ; q^{2}, a \rp
+ a^{+ \frac{1}{2}}
\frac{( a^{- 1} q, q^{3} ; q^{2} )_{\infty}}{( a^{- 1} q^{3}, q ; q^{2} )_{\infty}}
\ {}_{2}\varphi_{1} \lp a^{- 1} q^{2}, a^{- 1} q^{3} ; q^{3} ; q^{2}, a \rp
= \frac{( a^{- \frac{1}{2}} q ; q )_{\infty}}{( a^{+ \frac{1}{2}} ; q )_{\infty}}.
\label{amirror4}
\end{align}
One can easily conform the correctness of \eqref{amirror4} because this emerges on the way of the proof in the next subsection.

\paragraph{Generalized index} 
The generalized index is defined as the SCI with gauging flavor symmetries \cite{Kapustin:2011jm}. In our context, we introduce a background flat gauge field $B_{\rm flat}^{\rm flavor}$ by gauging the U$(1)_{A}$ flavor symmetry, and the parity conditions must be classified in terms of both the holonomy for the dynamical gauge field $A_{\rm flat}$ and the holonomy for $B_{\rm flat}^{\rm flavor}$, that is,
\begin{align}
e^{i \oint_{\g} ( \ch A_{\rm flat} + \fl B_{\rm flat}^{\rm flavor} )} = \pm 1,
\end{align}
as explained in Appendix \ref{Appmat}. Since our argument for theories with a single flavor does not change, our generalized indices are still \eqref{SQED1} and \eqref{xyz}, and mirror symmetry with gauging a flavor symmetry can also be concluded as \eqref{amirror3}. The generalized index carries much more important roles when one discusses the multiflavor case of mirror symmetry.

\subsection{Mathematical proof of $\mathbb{R}\mathbb{P}_{b}^{2} \times \mathbb{S}^{1}$}
In this subsection, we give the proof of our new relation \eqref{amirror3}. At first, we review the $q$-binomial theorem \cite{GR} derived mainly by Cauchy \cite{Cauchy} and Heine \cite{Heine},
\begin{equation}
{}_1\varphi_0(a;-;q,x)=\frac{(ax;q)_\infty}{(x;q)_\infty}, \quad |x|<1.
\label{qbinomial}
\end{equation}
This formula is the $q$-analogue of the \textit{binomial theorem}
\begin{align}
{}_2F_1(a,c;c;z)={}_1F_0(a;-;z)=\sum_{n\ge 0}\frac{(a)_n}{n!}z^n=(1-z)^{-a},
\end{align}
where $|z|<1$. $(a)_n$ is the classical shifted factorial $(a)_n=a(a+1)\dots (a+n-1)$, and $(a)_0=1$. We prove our new relation \eqref{amirror3} by utilizing the $q$-binomial theorem. The starting point is the SCI for the XYZ model \eqref{xyz},
\begin{align} 
\mathcal{I}_{\textrm{XYZ}}(x,\tilde{\alpha})
&=q^{+\frac{1}{8}}\tilde{a}^{-\frac{1}{4}}\frac{(\tilde{a}^{-\frac{1}{2}}q;q)_\infty}{(\tilde{a}^{+\frac{1}{2}};q)_\infty}\frac{(\tilde{a};q^2)_\infty}{(\tilde{a}^{-1}q;q^2)_\infty}\notag\\ 
&=q^{+\frac{1}{8}}\tilde{a}^{-\frac{1}{4}}\frac{(\tilde{a};q^2)_\infty}{(\tilde{a}^{-1}q;q^2)_\infty}\frac{(\tilde{a}^{-\frac{1}{2}}q;q)_\infty}{(\tilde{a}^{+\frac{1}{2}};q)_\infty}\notag\\ 
&=q^{+\frac{1}{8}}\tilde{a}^{-\frac{1}{4}}\frac{(\tilde{a};q^2)_\infty}{(\tilde{a}^{-1}q;q^2)_\infty}
{}_1\varphi_0(\tilde{a}^{-1}q;-;q,\tilde{a}^{+\frac{1}{2}})\notag\\ 
&=q^{+\frac{1}{8}}\tilde{a}^{-\frac{1}{4}}\frac{(\tilde{a};q^2)_\infty}{(\tilde{a}^{-1}q;q^2)_\infty}
\sum_{n\ge 0}\frac{(\tilde{a}^{-1}q;q)_n}{(q;q)_n}\left(\tilde{a}^{+\frac{1}{2}}\right)^n\notag\\ 
&=q^{+\frac{1}{8}}\tilde{a}^{-\frac{1}{4}}\frac{(\tilde{a};q^2)_\infty}{(\tilde{a}^{-1}q;q^2)_\infty}
\left\{\sum_{m\ge 0}\frac{(\tilde{a}^{-1}q;q)_{2m}}{(q;q)_{2m}}\left(\tilde{a}^{+\frac{1}{2}}\right)^{2m}
+\sum_{m\ge 0}\frac{(\tilde{a}^{-1}q;q)_{2m+1}}{(q;q)_{2m+1}}\left(\tilde{a}^{+\frac{1}{2}}\right)^{2m+1}\right\}. 
\label{alt1}
\end{align}
We remark that there are the following relations:
\begin{equation}\label{trs}
(a;q)_{2m}=(a,aq;q^2)_m, \quad (a;q)_{2m+1}=(1-a) (aq,aq^2;q^2)_m.
\end{equation}
We apply the relations \eqref{trs} to \eqref{alt1},
\begin{align} 
q^{+\frac{1}{8}}\tilde{a}^{-\frac{1}{4}}\frac{(\tilde{a};q^2)_\infty}{(\tilde{a}^{-1}q;q^2)_\infty}
&\left\{\sum_{m\ge 0}\frac{(\tilde{a}^{-1}q;q)_{2m}}{(q;q)_{2m}}\left(\tilde{a}^{+\frac{1}{2}}\right)^{2m}
+\sum_{m\ge 0}\frac{(\tilde{a}^{-1}q;q)_{2m+1}}{(q;q)_{2m+1}}\left(\tilde{a}^{+\frac{1}{2}}\right)^{2m+1}\right\}\notag\\ 
=q^{+\frac{1}{8}}\tilde{a}^{-\frac{1}{4}}\frac{(\tilde{a};q^2)_\infty}{(\tilde{a}^{-1}q;q^2)_\infty}
&\left\{\sum_{m\ge 0}\frac{(\tilde{a}^{-1}q,\tilde{a}^{-1}q^2;q)_m}{(q;q^2)_m(q^2;q^2)_m}\left(\tilde{a}\right)^m
+\tilde{a}^{+\frac{1}{2}}\frac{1-\tilde{a}^{-1}q}{1-q}
\sum_{m\ge 0}\frac{(\tilde{a}^{-1}q^2,\tilde{a}^{-1}q^3;q)_m}{(q^3;q^2)_m(q^2;q^2)_m}\left(\tilde{a}\right)^m
\right\}\notag\\ 
=q^{+\frac{1}{8}}\tilde{a}^{-\frac{1}{4}}\frac{(\tilde{a};q^2)_\infty}{(\tilde{a}^{-1}q;q^2)_\infty}
&\left\{{}_2\varphi_1(\tilde{a}^{-1}q,\tilde{a}^{-1}q^2;q;q^2,\tilde{a}) 
+\tilde{a}^{+\frac{1}{2}}\frac{1-\tilde{a}^{-1}q}{1-q}
{}_2\varphi_1(\tilde{a}^{-1}q^2,\tilde{a}^{-1}q^3;q^3;q^2,\tilde{a})
\right\}. 
\label{alt2}
\end{align}
The part $(1-\tilde{a}^{-1}q)/(1-q)$ can be rewritten as
\begin{equation}\label{trs2}
\frac{1-\tilde{a}^{-1}q}{1-q}=\frac{(\tilde{a}^{-1}q:q^2)_\infty}{(\tilde{a}^{-1}q^3;q^2)_\infty}\frac{(q^3;q^2)_\infty}{(q;q^2)_\infty}.
\end{equation}
Combining the relations \eqref{alt2} and \eqref{trs2}, we have
\begin{align} 
& q^{+\frac{1}{8}}\tilde{a}^{-\frac{1}{4}}\frac{(\tilde{a};q^2)_\infty}{(\tilde{a}^{-1}q;q^2)_\infty}
\left\{{}_2\varphi_1(\tilde{a}^{-1}q,\tilde{a}^{-1}q^2;q;q^2,\tilde{a}) \right.\notag\\
&\hspace{8em} + \left.\tilde{a}^{+\frac{1}{2}}\frac{1-\tilde{a}^{-1}q}{1-q}
{}_2\varphi_1(\tilde{a}^{-1}q^2,\tilde{a}^{-1}q^3;q^3;q^2,\tilde{a})
\right\} \non\\ 
=&\hspace{.25em} q^{+\frac{1}{8}}\frac{(\tilde{a};q^2)_\infty}{(\tilde{a}^{-1}q;q^2)_\infty}
\left\{\tilde{a}^{-\frac{1}{4}}{}_2\varphi_1(\tilde{a}^{-1}q,\tilde{a}^{-1}q^2;q;q^2,\tilde{a})\right. \non\\ 
&\hspace{6.65em} + \left. \tilde{a}^{+\frac{1}{4}}\frac{(\tilde{a}^{-1}q:q^2)_\infty}{(\tilde{a}^{-1}q^3;q^2)_\infty}\frac{(q^3;q^2)_\infty}{(q;q^2)_\infty}
{}_2\varphi_1(\tilde{a}^{-1}q^2,\tilde{a}^{-1}q^3;q^3;q^2,\tilde{a})\right\} \non\\ 
=&\hspace{.25em} q^{+\frac{1}{8}}\frac{(q^2;q^2)_\infty}{(q;q^2)_\infty}
\left\{\tilde{a}^{-\frac{1}{4}}\frac{(\tilde{a},q;q^2)_\infty}{(\tilde{a}^{-1}q,q^2;q^2)_\infty}{}_2\varphi_1(\tilde{a}^{-1}q,\tilde{a}^{-1}q^2;q;q^2,\tilde{a})\right. \non\\ 
&\hspace{5.65em} + \left. \tilde{a}^{+\frac{1}{4}}\frac{(\tilde{a},q^3;q^2)_\infty}{(\tilde{a}^{-1}q^3,q^2;q^2)_\infty}{}_2\varphi_1(\tilde{a}^{-1}q^2,\tilde{a}^{-1}q^3;q^3;q^2,\tilde{a})\right\}. 
\label{alt3}
\end{align}
Therefore, we obtain the conclusion
\begin{align} 
q^{+\frac{1}{8}}\tilde{a}^{-\frac{1}{4}}\frac{(\tilde{a}^{-\frac{1}{2}}q;q)_\infty}{(\tilde{a}^{+\frac{1}{2}};q)_\infty}\frac{(\tilde{a};q^2)_\infty}{(\tilde{a}^{-1}q;q^2)_\infty}
=q^{+\frac{1}{8}}\frac{(q^2;q^2)_\infty}{(q;q^2)_\infty}
&\left\{\tilde{a}^{-\frac{1}{4}}\frac{(\tilde{a},q;q^2)_\infty}{(\tilde{a}^{-1}q,q^2;q^2)_\infty}{}_2\varphi_1(\tilde{a}^{-1}q,\tilde{a}^{-1}q^2;q;q^2,\tilde{a})\right.\non \\ 
&\hspace{-.35em} + \left.\tilde{a}^{+\frac{1}{4}}\frac{(\tilde{a},q^3;q^2)_\infty}{(\tilde{a}^{-1}q^3,q^2;q^2)_\infty}{}_2\varphi_1(\tilde{a}^{-1}q^2,\tilde{a}^{-1}q^3;q^3;q^2,\tilde{a})\right\}. 
\label{final}
\end{align}

\section{Discussion} \label{Discussion}
We presented how to define $\mathcal{N}=2$ supersymmetric gauge theories on $\Rp_b^2 \times \Sp^1$ and got the exact form of the superconformal index with an arbitrary number of vector multiplets and matter multiplets with a U(1) gauge symmetry. As commented, the results are not dependent on $l$ and $\tilde{l}$, that is, the squashing parameter $b$. This fact is expected because it is verified in 2d cases \cite{Gomis:2012wy, Kim:2013ola}. Also, we gave the exact check of $\mathcal{N}=2$ and $\Ncal = 4$ Abelian mirror symmetry with the simplest case $N_f =1$ and Abelian duality (Appendix \ref{AD}) on $\Rp_b^2 \times \Sp^1$ by using the $q$-binomial theorem essentially. In the rest of this section, we would like to comment briefly on some open questions and future directions.

\paragraph{Open questions and future directions}
The first question is related to a subtlety in our computation of the superconformal index. We used an \textit{ad hoc} way to regulate the Casimir energy presented in \cite{Imamura:2011su, Imamura} (see Appendixes \ref{1-loop} and \ref{Formula}). They showed that the precise Chern-Simons level shift on $\Sp^2 \times \Sp^1$ emerges within this regularization scheme. However, as noted in Section \ref{SUSYbasis}, we cannot take the Chern-Simons term into account. Therefore, we cannot adopt the level shift as the guiding principle of the regularization and do not know why our regularization of the Casimir energy works so well. It is interesting to find more fundamental treatment to resolve it.
As the second one, we would like to know the origin of our B-parity condition on the XYZ model side. We took a little bit of an \textit{ad hoc} way to determine it based on the correspondence between the moduli spaces. In addition, we should check precisely whether our B-parity is unique or not. One straightforward way to solve this problem is using the brane construction of mirror symmetry \cite{deBoer:1997ka}. We expect that the generalized mathematical formulas will emerge if this program is accomplished.
The third question is related to the so-called ``factorization" property of 3d exact results \cite{Pasquetti:2011fj, Beem:2012mb, Hwang:2012jh, Taki:2013opa}. The partition functions on $\Sp^3_b$ and the superconformal indices on $\Sp^2 \times \Sp^1$ can be decomposed into the product of more fundamental quantities called \textit{holomorphic blocks}. This property of both cases naively comes from the fact that each curved space is characterized by solid-torus decomposition. However, $\Rp^2 \times \Sp^1$ cannot be expressed simply by using solid-torus decomposition. Instead, one can get $\Rp^2 \times \Sp^1$ by gluing the surface of a solid torus in an appropriate manner. One may find the unexpected description of our results in terms of holomorphic blocks via this method.
The final comment is concerned with an extension of our arguments. There are obvious open problems; we did not perform the check of Abelian mirror symmetry with general $N_{f}$ flavors. Also, we did not present the generalization of the exact calculation with a non-Abelian gauge symmetry as mentioned in the Introduction. We hope to complete these problems in the near future. Moreover, we found generalized mirror symmetry equalities in Appendix \ref{gene}. In Appendix \ref{qbin}, we provided the generalized equality with the parameter $\lambda$ and its proof in terms of the $q$-binomial theorem. In Appendix \ref{theta}, we showed another relation derived by the properties of the theta function of Jacobi. The idea of the proof comes from \textit{connection problems on linear $q$-difference equations} \cite{RSZ, Mbi}. The generalized relation also gives the \textit{connection formula} for the ${}_1\varphi_0(\lambda ;-;q,z)$-type equation between the solutions of the linear $q$-difference equations around the origin and around infinity. The important point is that we obtain the same relation \eqref{amirror3} as the special case even though these relations in the subsections are essentially different from each other. These formulas suggest the possibility to add one more parameter to our system, and its physical meaning may be found in the brane construction. If these are derived from string theory generally, our mathematical conclusion will give us new physical perspective.

\subsection*{Acknowledgments}
We would like to thank Heng-Yu Chen, Dongmin Gang, Kazuo Hosomichi, Yosuke Imamura, Yu Nakayama, Yousuke Ohyama, Satoshi Yamaguchi, and Yutaka Yoshida. We are grateful to the Yukawa Institute for Theoretical Physics at Kyoto University for hospitality during the workshop YITP-W-14-4 ``Strings and Fields," where part of this work was carried out. The work of A.T., H.M., and T.M. was supported in part by the JSPS Research Fellowship for Young Scientists.

\appendix
\section{Calculation details}\label{1-loop}
In this appendix, we show the details of the calculations for one-loop determinants. Our method discussed below is similar to the one discussed in \cite{Gomis:2012wy, Kim:2013ola}. Their way did not respect the symmetry generated by $\hat{j}_3$, whereas we derive \eqref{vec1loop} with preserving $\hat{j}_3$ structure explicitly because it has an important meaning in our SCI \eqref{indexdef}. In the latter discussions, we get the following type of a infinite product in each final step:
\begin{align}
\prod_{n\in \mathbb{Z}}
\prod_{k \geq 0}
\frac
{ 2\pi i n + 2z_f(k)}
{ 2\pi i n +  2z_b(k)}
,
\label{genericform}
\end{align}
where the $z_{f/b}(k)$'s represent certain $k$-dependent functions. By using the infinite product formula of $\sinh z$, we can deform it to
\begin{align}
&
\prod_{k \geq 0}
\frac
{2 \sinh z_f (k)}
{2 \sinh z_b (k)}
=
\Big(
\prod_{k \geq 0}
e^{z_f(k) - z_b(k)}
\Big)
\times
\exp{\Big(
\sum_{m=1}^\infty
\frac{-1}{m}
\sum_{k \geq 0}
(e^{-2mz_f(k)}
-
e^{-2mz_b(k)})
\Big)}.
\label{generic2}
\end{align}
We call the first part in \eqref{generic2} the \textit{Casimir energy} which must be regularized (see Appendix \ref{Formula} for our regularization scheme) and the second part,
\begin{align}
-
\sum_{k \geq 0}
(e^{-2z_f(k)}
-
e^{-2z_b(k)})
=:
f (x, \dots),
\end{align}
the \textit{one-particle index}.
As one can verify later, both the Casimir energy and the one-particle index do not depend on $x'$.
In this appendix, we use 2d Killing spinors
\begin{align}
\varepsilon(\compe, \compt)
=
e^{\frac{i}{2}\compt}
\begin{pmatrix}
\cos \frac{\compe}{2} \\
\sin \frac{\compe}{2}
\end{pmatrix},
\quad
\overline{\varepsilon}(\compe, \compt)
=
e^{-\frac{i}{2}\compt}
\begin{pmatrix}
\sin \frac{\compe}{2} \\
\cos \frac{\compe}{2}
\end{pmatrix}
\end{align}
which satisfy
\begin{align}
\D_i \varepsilon = \frac{1}{2f} \g_i \g_3 \varepsilon,
\quad
\D_i \overline{\varepsilon} = -\frac{1}{2f} \g_i \g_3 \overline{\varepsilon},
\end{align}
where $i$ runs for $\compe, \compt$.
We must consider the B-parity condition in order to get the index on $\Rp_b^2 \times \Sp^1$.
If we ignore the B-parity condition, then, of course, we can get the index on $\Sp_b^2 \times \Sp^1$.
However, as noted in the beginning of Section \ref{SUSYbasis}, the results do not depend on the squashing parameter.
Consequently, the results without the B-parity condition reproduce the known results on $\Sp^2 \times \Sp^1$ \cite{Kim:2009wb, Gang:2009wy, Imamura:2011su}.

\subsection{Vector multiplet}\label{Appvec}
\paragraph{Gauge fixing}
By repeating the same argument for the ``shortcut" way of the gauge fixing \cite{Hama:2011ea, Gomis:2012wy, Kim:2013ola}, we can restrict the path integral onto the configuration satisfying
\begin{align}
A_3^{(n)} &= 0,
\label{A3zero}
\\
*_2 d *_2 \A^{(n)} &=0
\label{2dgaugefix}
\end{align}
for all $n$'s without any Fadeev-Popov determinants. Then, we need to consider the operator's determinant
\begin{align}
& \prod_{n \in \mathbb{Z}} \frac{\det \Delta_f^{(n)}}{\sqrt{\det \Delta_b^{(n)}}},
\end{align}
where
\begin{align}
&\Delta_b^{(n)}
=
\begin{pmatrix}
- (*_2d)^2 + \h^2 & - *_2d \frac{1}{f} \\
+\frac{1}{f} *_2 d  & - (*_2 d)^2 + \frac{1}{f^2} + \h^2 
\end{pmatrix}
,
\\
&\Delta_f^{(n)}
=
i \g^i \D_i
+  \g_3 \big( \h + \frac{i}{2 l} \tri \big)
.
\end{align}
In addition, we can make this problem simpler by notifying 
\begin{align}
\det \dl_b^{(n)} = \sqrt{ \det \Delta_b^{(n)}}
\end{align}
up to the sign where
\begin{align}
\dl_b^{(n)} 
=
\begin{pmatrix}
 i \h & - *_2d \\
*_2d & \frac{1}{f} + i \h 
\end{pmatrix}
.
\end{align}
Namely, the one-loop determinant, which we should know\footnote{The insertion of $\g_3$ in the numerator does not spoil the validity and make the problem simple \cite{Gomis:2012wy, Kim:2013ola}.} is
\begin{align}
&
\mathcal{Z}_{\rm 1-loop}^{\text{vector }}
=
 \prod_{n \in \mathbb{Z}} \frac{\det\g_3 \Delta_f^{(n)}}{\det \delta_b^{(n)}}
.
\label{vecoloop}
\end{align}
As one can see, the contribution of the U$(1)$ vector multiplet already does not have the dependence on the holonomy. Therefore, we omit the superscript $\pm$ from now on.

\paragraph{Pairing structure}
The calculation is based on the eigenvalues pairing structure as follows. Let $(\A, \si)^\text{T}$ and $\lam$ be the eigenmodes,
\begin{align}
\dl_{b}^{(n)} 
\begin{pmatrix}
\A \\
\si
\end{pmatrix}
 &= -i M 
 \begin{pmatrix}
\A \\
\si
\end{pmatrix}
 ,
\label{vecbe}
\\
\Delta_{f}^{(n)} \lam &= - M \lam.
\label{vecfe}
\end{align}
Then, we can map the one side to the other by defining
\begin{align}
\Lam &:= (\g_3 \g^i \A_i + i  \si \g_3) \varepsilon,
\label{pairLam}
\\
\begin{pmatrix}
\mathcal{B} \\
\Sigma
\end{pmatrix}
&:=
\begin{pmatrix}
 -i (M + \h) \overline{\varepsilon} \g_i \lam e^i - d (\overline{\varepsilon} \g_3 \lam) \\
(M + \h )  \overline{\varepsilon} \lam
\end{pmatrix}.
\label{pairB}
\end{align}
The modes which have no pair only contribute to the one-loop determinant \eqref{vecoloop}.
In other words, we have to find the eigenvalues constrained by the following conditions:
\begin{align}
&M=M_b
\text{ which satisfies }
\eqref{vecbe}
\text{ and }
\eqref{pairLam}=0,
\label{relvbos}
\\
&M=M_f
\text{ which satisfies }
\eqref{vecfe}
\text{ and }
\eqref{pairB}=0.
\label{relvfer}
\end{align}
The constraints $\eqref{pairLam}=0$ and $\eqref{pairB}=0$ are solved by taking
\begin{align}
\begin{pmatrix}
\A \\
\si
\end{pmatrix}
&=
e^{i j_3^b \compt}
 h_b(\compe) 
\begin{pmatrix}
e^1 + i \cos \compe e^2 \\
i \sin \compe
\end{pmatrix},
\label{constvb}
\\
\lam &= 
(M_f + \h + \frac{i}{2 l} \tri)e^{i j_3^f \compt}
 h_f(\compe)   \overline{\varepsilon} ,
\label{constvf}
\end{align}
where $j_3^{b/f} \in \mathbb{Z}$. Substituting these representations into \eqref{vecbe} and \eqref{vecfe}, we get the following sets of equations:
\begin{align}
&
\left\{ \begin{aligned}
&\frac{1}{f(\compe)} \pa_\compe h_b(\compe)
+ \frac{\cos \compe}{\sin \compe } \Big(  \frac{1}{f(\compe)} - \frac{j_3^b}{l} \Big) h_b(\compe)
=0, \\ 
&M_bl
=
i\Big(
(\tri -1) j_3^b
-  i l \darn
\Big), 
\end{aligned} \right.
\label{vrpb}
\\
&
\left\{ \begin{aligned}
&\frac{1}{f(\compe)}  \pa_\compe h_f(\compe)
+
\frac{\cos \compe}{\sin \compe} \Big( \frac{1}{f(\compe)} 
+\frac{ j_3^f -1}{l}
  \Big)  
h_f(\compe)
=0, \\ 
&M_f l =  i \Big( (\tri - 1)( j_3^f -1) - i l \darn   \Big). 
\end{aligned} \right.
\label{vrpf}
\end{align}
One can get the conditions of $j_3$ as $j_3^b \geq 1$ for bosons and $j_3^f \leq 0$ for fermions because, around $\compe \sim 0$, one can easily solve the equation for $h_b (\compe)$ and $h_f(\compe)$ in \eqref{vrpb} and \eqref{vrpf}, respectively, as
\begin{align}
&h_b(\compe) \sim \sin^{(j_3^b-1)} \compe,
\\
&h_f(\compe) \sim \sin^{-j_3^f} \compe.
\end{align}
Note that the coefficients of differential equations with respect to $\compe$ in \eqref{vrpb} and \eqref{vrpf} are invariant under the antipodal identification \eqref{antipodal2}, that is,
\begin{align}
h_{b / f} (\pi - \compe)
=
h_{b / f} (\compe).
\label{hinvvec}
\end{align}

\paragraph{B-parity condition}
Usually, $j_3$ takes an arbitrary value in integers $\mathbb{Z}$. Therefore, one may think that $j_3^b = 1,2,3,...$ and $j_3^f = 0, -1, -2,...$; however, it is \textit{not} in our case because of the B-parity condition. We can determine the possible values for $j_3^{b/f}$ from the explicit forms of the eigenmodes \eqref{constvb} and \eqref{constvf}, the invariance of $h_{b/f}$ \eqref{hinvvec}, and the B-parity condition \eqref{vecp}. Combining these arguments, one can get the condition
\begin{align}
e^{ij_3 \pi} &= -1.
\end{align}
This means that we have 
\begin{align}
\left\{ \begin{aligned}
M_b &= \frac{i}{l} \Big( (\tri -1)(2k +1) - i l \darn \Big), \\
M_f &= \frac{i}{l} \Big( -(\tri -1)(2k +2) - i l \darn \Big),
\end{aligned} \right.
\label{eigenvalues}
\end{align}
where $k=0,1,2, ...$, and $n \in \mathbb{Z}$. Note that the eigenvalues for bosons shift by one from those of fermions, which results in the nontrivial one-loop determinant for the U$(1)$ vector multiplet.

\paragraph{One-loop determinant}
We can get the explicit form of \eqref{vecoloop} just by substituting all relevant eigenvalues \eqref{eigenvalues} into it:
\begin{align}
\eqref{vecoloop}
&=
\prod_{\rm all} \frac{M_f}{M_b}
\notag \\
&=
\prod_{n \in \mathbb{Z}}
\Big(
\prod_{k \geq 0}
\frac{(1-\tri)(2k +2) - i l \darn}{(\tri - 1)(2k+1) - i l \darn}
\Big)
\notag \\
&\sim
\prod_{n \in \mathbb{Z}}
\Big(
\prod_{k \geq 0}
\frac{(1-\tri)(2k +2) + i l \darn}{(1-\tri)(2k+1) + i l \darn}
\Big),
\label{vecoloop2}
\end{align}
where $\sim$ represents the equality up to the sign. This regularization is guaranteed in the 2d case \cite{Kim:2013ola}. From the above expression, substituting
\begin{align}
&2z_f(k)=
 2 \beta_2 (2k+2 )
,
\qquad
2z_b(k)
=
 2 \beta_2 (2k+1 )
\label{vecCasimir}
\end{align}
into \eqref{generic2}, we can get \eqref{vec1loop} and \eqref{vecind}. The Casimir energy can be regularized by using the zeta function regularization formula \eqref{Casimirreg} explained in Appendix \ref{Formula}.

\subsection{Matter multiplet}\label{Appmat}
We start with the pairing structure of \eqref{matbosd} and \eqref{matferd}. To make our argument comprehensive, we define the differential operators
$\Delta_\phi^{(n)}$ and $\Delta_\psi^{(n)}$
acting on $\phi^{(n)}$ and $\psi^{(n)}$, respectively, as
\begin{align}
&\Delta_\phi^{(n)}
=
- g^{i j} \D_i^{A_\fla} \D_j^{A_\fla}
+(\p - i \frac{\Dl}{2 l}\tri )^2 
+ \frac{\Delta^2 - 2 \Delta}{4f^2}
+ \frac{\Delta}{4}R
-\frac{\Delta-1}{f}v^i   \D_i^{A_\fla}
,
 \label{matbosdap}
 \\
&\Delta_\psi^{(n)}
=
-i  \g^i \D_i^{A_\fla}
- \g_3 (\p - i \frac{\Delta-1}{2 l} \tri )
- i \frac{1}{2f} \g_3 
-i\frac{\Delta-1}{2f}v^i  \g_i 
-i\frac{\Delta-1}{2f} \om 
,
 \label{matferdap}
\end{align}
where $\Dcal_{i}^{A_{\rm flat}}$ is defined with a flat connection $A_{\rm flat}$.

\paragraph{Pairing structure}
Let $\phi$ and $\psi$ be the eigenmodes for $\Delta_\phi^{(n)}$ and $\Delta_\psi^{(n)}$, i.e.
\begin{align}
&\Delta_\phi^{(n)} \phi = - M \Big(M - 2 ( \p - i \frac{\Dl}{2 l} \tri ) \Big) \phi,
\label{matbe}
\\
&\g_3 \Delta_f^{(n)}\psi = M\psi.
\label{matfe}
\end{align}
Then,
\begin{align}
\begin{pmatrix}
\Psi_1 \\
\Psi_2
\end{pmatrix}
&=
\begin{pmatrix}
\g_3 \varepsilon \phi \\
 i \g^i \varepsilon \D_i^{A_\fla} \phi 
 + \g^3 \varepsilon
 ( [\h- i \frac{\Dl}{2 l} \tri ]+ i \frac{\Delta}{2f})
 \phi
\end{pmatrix},
\label{pairPsi}
\\
\Phi &= \overline{\varepsilon} \psi
\label{pairPhi}
\end{align}
satisfy the equations
\begin{align}
&
\g_3 \Delta_\psi^{(n)}
\begin{pmatrix}
\Psi_1 \\
\Psi_2
\end{pmatrix}
=
\begin{pmatrix}
-2 (\p - i \frac{\Dl - 1}{2 l} \tri ) & 1 \\
M(M- 2 (\p - i \frac{\Dl - 1}{2 l} \tri )) & 0
\end{pmatrix}
\begin{pmatrix}
\Psi_1 \\
\Psi_2
\end{pmatrix}
,
\label{pairedfe}
\\
&
\Delta_\phi^{(n)} \Phi = - M \Big(M - 2 ( \p - i \frac{\Dl}{2 l} \tri ) \Big) \Phi.
\label{pairedbe}
\end{align}
As discussed in \cite{Hama:2011ea, Gomis:2012wy, Kim:2013ola}, one can find the relevant spectra characterized by
\begin{align}
&M=M_\phi \text{ which satisfies } \eqref{matbe} \text{ and }\Psi_2 = M \Psi_1,
\label{relmbos}
\\
&M=M_\psi \text{ which satisfies } \eqref{matfe} \text{ and }\eqref{pairPhi}=0.
\label{relmfer}
\end{align}
Then, we take each relevant mode as
\begin{align}
&\phi 
= 
e^{i \oint_{\g} \ch A_\fla} e^{i j_3^b \compt}
h_b(\compe),
\label{constmb}
\\
&\psi =
e^{i \oint_{\g} \ch A_\fla} e^{i j_3^f \compt}
h_f(\compe) \overline{\varepsilon},
\label{constmf}
\end{align}
where $j_3^{b/f} \in \mathbb{Z}$. Substituting these forms into $\Psi_2 = M_\phi \Psi_1$ and \eqref{pairPhi} $=0$, we get the following sets of equations:
\begin{align}
&
\left\{ \begin{aligned}
& \frac{1}{f}  \pa_\compe  h_b (\compe)
 +
\frac{\cos \compe}{\sin \compe}
\Big( \frac{\Delta}{2f(\compe)} 
-
\frac{1}{l} ( j_3^b + \frac{\Delta}{2} )\Big)h_b(\compe)
=0
, \\ 
& M_\phi l
=
i
\Big(
(1-\tri)
(j_3^b+ \frac{\Delta}{2})
+i \frac{l}{2 \pi \rad} [2 \pi n - \ch \theta - \fl\mu]
\Big), 
\end{aligned} \right.
\label{vsb}
\\
&
\left\{ \begin{aligned}
& \frac{1}{f}  \pa_\compe  h_f (\compe)
-
\frac{\cos \compe}{\sin \compe}
\Big(
 \frac{\Delta-2}{2f(\compe)}
-\frac{1}{l} (j_3^f +  \frac{\Delta-2}{2})
 \Big)  h_f(\compe)=0,
 \\ 
& M_\psi l
=
i \Big(
(\tri - 1)
(j_3^f + \frac{\Delta -2}{2})
-i\frac{l}{2\pi \rad}[2 \pi n - \ch \theta- \fl \mu]
 \Big). 
\end{aligned} \right.
\label{vsf}
\end{align}
One can get also the conditions of $j_3$ as $j_3^b \geq 0$ for bosons and $j_3^f \leq 0$ for fermions because the behaviors of $h_{b/f} ( \vartheta )$ around $\compe \sim 0$ become
\begin{align}
&h_b(\compe) \sim \sin^{j_3^b} \compe,
\\
&h_f (\compe) \sim \sin^{-j_3^f} \compe.
\end{align}
Note that these functions Eqs. \eqref{vsb} and \eqref{vsf} have the symmetry \eqref{hinvvec}.

\paragraph{B-parity condition}
We have to limit $j_3$ to preserve the B-parity conditon \eqref{ematp} as we have done in the vector multiplet, but an additional issue occurs because the matter is charged through $\ch$. The permitted region depends on the B-parity choice $\pm$ and the value of the holonomy,
\begin{align}
e^{i \oint_\g \ch A_\fla} e^{ij_3 \pi}
=
\pm1.
\end{align}
Here, it is found that the consistent two choices of the B-parity condition correspond to the background U$(1)_{\text{flavor}}$ holonomies
\begin{align}
\pm1 = e^{i \oint_\g \fl B_\fla^{\text{flavor}}},
\end{align}
where $\fl$ is the appropriate flavor charge.
Therefore, we can get
\begin{align}
e^{ij_3 \pi}
=
e^{i \oint_\g (\ch A_\fla +  \fl B_\fla^{\text{flavor}})} 
.
\end{align}
It means that we have
\begin{align}
\Bigg(
e^{i \oint_\g (\ch A_\fla +  \fl  B_\fla^{\text{flavor}})} =+1
\Bigg)
&\Rightarrow
\left\{ \begin{aligned}
M_\phi l
&=
i
\Big(
(1-\tri)
(2k+ \frac{\Delta}{2})
+i \frac{l}{2 \pi \rad}
[2 \pi n - \ch\theta -\fl \mu]
\Big),
\\ 
M_\psi l
&=
i \Big(
(\tri - 1)
(-2k-1 + \frac{\Delta }{2})
-i\frac{l}{2\pi \rad}[2 \pi n - \ch \theta - \fl \mu]
 \Big), 
\end{aligned} \right.
\label{eigenvalues1}
\\
\Bigg(
e^{i \oint_\g (\ch A_\fla +  \fl B_\fla^{\text{flavor}})} =-1
\Bigg)
&\Rightarrow
\left\{ \begin{aligned}
M_\phi l
&=
i\Big((1-\tri)(2k+1+ \frac{\Delta}{2})+i \frac{l}{2 \pi \rad}[2 \pi n - \ch \theta-\fl \mu]\Big),
\\ 
M_\psi l
&=
i \Big(
(\tri - 1)
(-2k-2 + \frac{\Delta }{2})
-i\frac{l}{2\pi \rad}[2 \pi n - \ch\theta-\fl\mu]
 \Big). 
\end{aligned} \right.
\label{eigenvalues2}
\end{align}
Therefore, the one-loop determinant changes its form depending on the value of the total holonomy $e^{i \oint_\g (\ch A_\fla +  \fl  B_\fla^{\text{flavor}})}$.

\paragraph{One-loop determinant}
We can get each one-loop determinant by calculating
\begin{align}
\prod_{\rm all}
\frac{M_\psi}{M_\phi}
.
\label{matter1loop}
\end{align}
We read the eigenvalue of each holonomy sector from \eqref{eigenvalues1} and \eqref{eigenvalues2}, and the corresponding infinite products \eqref{matter1loop} are written as
\begin{align}
\Bigg(
e^{i \oint_\g (\ch A_\fla +  \fl  B_\fla^{\text{flavor}})} =+1
\Bigg)
&\Rightarrow
\Bigg(
\eqref{matter1loop}
=
\prod_{n \in \mathbb{Z}}
\prod_{k \geq 0}
\frac
{ (1-\tri)(2k+1 - \frac{\Delta }{2})-i\frac{l}{2\pi \rad}[2 \pi n - \ch \theta-\fl\mu] }
{(1-\tri)(2k+ \frac{\Delta}{2})+i \frac{l}{2 \pi \rad}[2 \pi n - \ch \theta-\fl\mu]}
\Bigg),
\label{matter1loop1}\\ 
\Bigg(
e^{i \oint_\g ( \ch A_\fla +  \fl  B_\fla^{\text{flavor}})} =-1
\Bigg)
&\Rightarrow
\Bigg(
\eqref{matter1loop}
=
\prod_{n \in \mathbb{Z}}
\prod_{k \geq 0}
\frac
{ (1-\tri)(2k+2 - \frac{\Delta }{2})-i\frac{l}{2\pi \rad}[2 \pi n - \ch \theta-\fl\mu] }
{(1-\tri)(2k+1+ \frac{\Delta}{2})+i \frac{l}{2 \pi \rad}[2 \pi n - \ch \theta-\fl\mu]}
\Bigg). 
\label{matter1loop2}
\end{align}
After substituting
\begin{align}
\Bigg(
e^{i \oint_\g (\ch A_\fla +  \fl B_\fla^{\text{flavor}})} =+1
\Bigg)
&\Rightarrow
\left( \begin{aligned}
2z_f (k)
&=
i (\ch\theta +\fl\mu ) + 2 \beta_2 (2k+1 - \frac{\Delta}{2} ),
\\ 
2z_b(k)
&=
-i ( \ch \theta +\fl\mu ) + 2 \beta_2 (2k + \frac{\Delta}{2} ),
\end{aligned} \right), \label{hompCasimir} 
\\ %
\Bigg(
e^{i \oint_\g (\ch A_\fla +  \fl B_\fla^{\text{flavor}})} =-1
\Bigg)
&\Rightarrow
\left( \begin{aligned}
2z_f(k)
&=
i (\ch \theta +\fl\mu ) + 2 \beta_2 (2k+2 - \frac{\Delta}{2} ),
\\ 
2z_b(k)
&=
-i (\ch \theta +\fl\mu ) + 2 \beta_2 (2k+1 + \frac{\Delta}{2} )
\end{aligned} \right) \label{hommCasimir} 
\end{align}
into \eqref{generic2} 
and regularizing the Casimir energies by using \eqref{Casimirreg},
we can get the results \eqref{mat+1loop} - \eqref{mat-ind}.

\section{Zeta function regularization} \label{Formula}
In general, an infinite product is not well defined and must be regulated by an appropriate method. Here, we adopt the zeta function regularization given as \cite{Imamura}
\begin{align}
\prod_{k \geq 0} f ( k ) = \exp \left. \lp \frac{d}{d s} \sum_{k \geq 0} f ( k )^{s} \rp \right|_{s = 0}.
\label{zetareg}
\end{align}
We make use of \eqref{zetareg} to regularize the Casimir energy of the vector multiplet and the matter multiplet. Those forms shown explicitly in \eqref{generic2} are generally written as the infinite product
\begin{align}
\prod_{k \geq 0} \frac{\lp x^{2 k + C_{1}} \rp^{r}}{\lp x^{2 k + C_{2}} \rp^{r}},
\label{casiform}
\end{align}
where $C_{1}, C_{2}$, and $r$ are constants independent of $k$. Applying \eqref{zetareg} to the above expression, we expand the numerator and the denominator around $s = 0$ so that 
\begin{align}
\frac{d}{d s} \sum_{k \geq 0} \lp x^{2 k + C} \rp^{s}
= \frac{1}{2 s^{2} \log x} + \frac{- 2 + 6 C - 3 C^{2}}{12} \log x + O ( s ).
\label{zetareg2}
\end{align}
Although this form is obliviously diverged at $s = 0$, unwanted terms can be canceled by taking a ratio of such infinite products. Consequently, \eqref{casiform} with $s \to 0$ results in
\begin{align}
\prod_{k \geq 0} \frac{\lp x^{2 k + C_{1}} \rp^{r}}{\lp x^{2 k + C_{2}} \rp^{r}}
&= \exp \lp r \lp \frac{- 2 + 6 C_{1} - 3 C_{1}^{2}}{12} - \frac{- 2 + 6 C_{2} - 3 C_{2}^{2}}{12} \rp \log x \rp \non \\ 
&= x ^{- \frac{r}{4} \lp C_{1} - C_{2} \rp \lp C_{1} + C_{2} - 2 \rp}.
\label{Casimirreg}
\end{align}
It is straightforward to apply this formula to each Casimir energy. Firstly, for the vector multiplet, its $k$-dependent functions \eqref{vecCasimir} correspond to setting
\begin{align}
r = 1, \hspace{1em}
C_{1} = 1, \hspace{1em}
C_{2} = 2.
\end{align}
Then, we can obtain the Casimir energy as
\begin{align}
x^{+ \frac{1}{4}}.
\label{casivv}
\end{align}
Secondly, for the matter multiplet in the even holonomy sector, its $k$-dependent functions \eqref{hompCasimir} correspond to setting
\begin{align}
r = 1, \hspace{1em}
C_{1} = \frac{\Dl}{2} + \Theta, \hspace{1em}
C_{2} = 1 - \frac{\Dl}{2} - \Theta, \hspace{1em}
x^{\Theta} := ( e^{+ i \ch \theta} \alpha^{+ \fl} )^{\frac{1}{2}}.
\end{align}
Then, we can obtain the Casimir energy as
\begin{align}
x^{+ \frac{\Dl - 1}{4}} e^{+ \frac{i}{4} \ch \theta} \alpha^{+ \frac{1}{4} \fl}.
\label{casip}
\end{align}
Lastly, for the matter multiplet in the odd holonomy sector, its $k$-dependent functions \eqref{hommCasimir} correspond to setting
\begin{align}
r = 1, \hspace{1em}
C_{1} = 1 + \frac{\Dl}{2} + \Theta, \hspace{1em}
C_{2} = 2 - \frac{\Dl}{2} - \Theta.
\end{align}
Then, we can obtain the Casimir energy as
\begin{align}
x^{- \frac{\Dl - 1}{4}} e^{- \frac{i}{4} \ch \theta} \alpha^{- \frac{1}{4} \fl}.
\label{casim}
\end{align}

\section{Abelian duality} \label{AD}
As recently discussed in \cite{Beasley:2014ila} for the purely bosonic case, though Abelian duality looks trivial on a flat space, its validity becomes nontrivial on the curved space because of topological obstructions. In this section, we utilize Abelian duality to justify our prescription (mainly, of the integration contour in the index).

Abelian duality in 3d between the free U$(1)$ gauge theory and the free matter theory \cite{Prodanov:1999jy, Broda:2002wg} can be realized by the equality of the action, 
\begin{align}
\int d^3 x \sqrt{g} \Big( \frac{1}{2} F_{\mu \nu} F^{\mu \nu} \Big)
=
\int d^3 x \sqrt{g} \Big(  \pa_\mu \rho \pa^\mu \rho \Big)
\label{dualityaction}
\end{align}
via the equality in \eqref{dualphoton}. We naturally can supersymmetrize this duality. For example, see \cite{Okazaki:2013kaa}. In our language, we can describe a (on-shell) matter multiplet $( \phi, \bar{\phi}, \psi, \bar{\psi} )$ in terms of a vector multiplet $( A_{\mu}( \to \rho ), \sg, \la, \bar{\la} )$ as follows:
\begin{align}
e^{2} \phi &= \sg + i \rho, \hspace{2em}
e^{2} \bar{\phi} = \sg - i \rho, \label{eq:ab1} \\
e^{2} \psi &= \la, \hspace{4em}
e^{2} \bar{\psi} = - \bar{\la}. \label{eq:ab3}
\end{align}
Through those identifications, one can show the following relationship between the U$(1)$ vector multiplet action and the matter multiplet action\footnote{Note that the duality equation \eqref{dualphoton} is valid on a Minkowski background. In order to make it Euclidean one, we have to multiply $(- i)$ by the third coordinate.}:
\begin{align}
& \frac{1}{e^{2}} \int d^3 x \sqrt{g} \left( \frac{1}{2} F_{\mu \nu}  F^{\mu \nu} + \pa_\mu \si  \pa^\mu \si + \e^{3 \rho \si} \frac{\si}{f} F_{\rho \si} 
+ i \olam \g^\mu \D_\mu \lam
- \frac{i}{2f} \olam \g_3 \lam \right) \notag \\ 
&= e^{2} \int d^3 x \sqrt{g} \left(
\pa_\mu \ophi \pa^\mu \phi  
+ \frac{1}{f} \ophi \pa_3 \phi
- i \opsi \g^\mu \D_\mu\psi
+ \frac{i}{2f} \opsi \g_3 \psi \right). 
\label{dmonshell}
\end{align}
The matter action above seems to be the on-shell part of $\Lcal_{\rm mat}$ with zero R-charge, $\Delta=0$. In fact, the dual matter fields under the identifications \eqref{eq:ab1} and \eqref{eq:ab3} can correctly reproduce the boundary conditions along $\mathbb{S}^1$ \eqref{kkbcm1} - \eqref{kkbcm4} setting $\Delta=0$. Moreover, the consistent B-parity condition for the vector multiplet \eqref{vecp} takes the one for the dual matter contents to be with negative sign in \eqref{ematp} representing the odd holonomy. Thus, we conclude that the matter multiplet comprised by the vector multiplet\footnote{Note that, as explained in \cite{Beasley:2014ila}, the invariance of the classical action \eqref{dualityaction} and the relation \eqref{dualphoton} under scale transformations requires the coupling constant $e$ to scale nontrivially such that $e^2$ has a scaling dimension one. This means that $\sg$ and $\rho$ in \eqref{eq:ab1} have effectively the same scaling dimension as that of the square of the coupling constant.} belongs to the odd holonomy sector with $\Delta=0$.

While the dual prescription shown here holds on the on-shell fields, we could reconstruct an off-shell action with $\Dl = 0$ and the supersymmetry for the dual matters by adding auxiliary fields appropriately to on-shell quantities (as the Gaussian form in the action).  Actually, the action \eqref{dmonshell} is just the on-shell sector having $\Delta=0$ of  the off-shell action
\begin{align}
\mathcal{L}_{\rm mat} + \dl_{\oep} \dl_{\e} \left( i \frac{\Delta-1}{f} [\ophi \phi] \right).
\label{dmatteraction}
\end{align}
As a result, because the matter action \eqref{dmatteraction} is written by a SUSY-exact deformation, we can perform the localization leading to the one-loop determinant for the corresponding dual matter fields. Therefore, the expected identity for Abelian duality is
\begin{align}
\mathcal{I}_{{\rm U}(1)}(x)
=
\mathcal{I}_{\text{matter}}^{(\Delta=0)(-)}(x)
.
\end{align}
The left-hand side can be computed by the formula \eqref{gSCIr}. Since we have just one vector multiplet, the contributions from the holonomies $e^{i \oint_\g A_\fla^{(\pm)}} = \pm1$ become the same ones:
\begin{align}
\mathcal{Z}_{\text{1-loop}}^{\text{vector}}
(x)
=
x^{\frac{1}{4}} \frac{(x^4 ; x^4)_\infty}{(x^2 ; x^4)_\infty}
.
\end{align}
This means that there is a residual symmetry of interchanging the flat connection for the even holonomy and for the odd holonomy $A_\fla^{(+)} \leftrightarrow A_\fla^{(-)}$, and we must take $(Sym) =2$ in the formula \eqref{gSCIr}:
\begin{align}
\mathcal{I}_{{\rm U}(1)}(x)
&=
\frac{\mathcal{Z}_{\text{1-loop}}^{\text{vector}}
(x)}{2} 
\oint_{C_0} \frac{d z}{2 \pi i z}
\Big( 
1
+
1
\Big)
\notag \\
&=
x^{\frac{1}{4}} \frac{(x^4 ; x^4)_\infty}{(x^2 ; x^4)_\infty}.
\label{indU1}
\end{align}
On the other hand, we can get the contribution from the dual matter multiplet via the formula in \eqref{nogSCIr} with $\Delta =\ch =\fl =0, N_f^{(+)} = N_f^{(2)}=0,$ and $N_f^{(-)}=1$:
\begin{align}
\mathcal{I}_{\text{matter}}^{(\Delta=0)(-)}(x)
&=
\mathcal{Z}_{\text{1-loop}}^{(\Delta=0)(-)} (1,x,1)
\notag \\
&=
x^{\frac{1}{4}} \frac{(x^4 ; x^4)_\infty}{(x^2 ; x^4)_\infty},
\end{align}
and this is identical to the one in \eqref{indU1}. Our new formulas to the index on $\mathbb{RP}_{b}^{2} \times \mathbb{S}^{1}$ can precisely provide Abelian duality as well as 3d mirror symmetry.

\section{Mathematical generalizations of \eqref{amirror3}} \label{gene}
In this section, we consider mathematical generalizations of the relation \eqref{amirror3}. In Appendix \ref{qbin}, we give \eqref{amirror3} as the special case of the generalization via the $q$-binomial theorem. In Appendix \ref{theta}, we also give \eqref{amirror3} by using the connection formula of ${}_1\varphi_0(\lambda ;-;q,z)$. We remark that these formulas in each subsection are completely different, but we can derive the relation \eqref{amirror3} as their special case.

\subsection{From the $q$-binomial theorem}\label{qbin}
First, we derive a more general form of \eqref{amirror3} from the $q$-binomial theorem and its \textit{alternative representation}. The $q$-binomial theorem is
\begin{align*}
\frac{(\lambda z;q)_\infty}{(z;q)_\infty}={}_1\varphi_0(\lambda ;-;q,z), \quad\forall |z|<1, |q|<1,
\end{align*}
and ${}_1\varphi_0(\lambda ;-;q,z)$ can be deformed from its definition as
\begin{align*}
{}_1\varphi_0(\lambda ;-;q,z)
&=\sum_{n\ge 0}\frac{(\lambda ;q)_n}{(q;q)_n}z^n \\ 
&=\sum_{m\ge 0}\frac{(\lambda ;q)_{2m}}{(q;q)_{2m}}z^{2m}
+\sum_{m\ge 0}\frac{(\lambda ;q)_{2m+1}}{(q;q)_{2m+1}}z^{2m+1}\\ 
&=\sum_{m\ge 0}\frac{(\lambda ,\lambda q;q^2)_{m}}{(q;q^2)_{m}(q^2;q^2)_m}(z^2)^{m}
+\frac{1-\lambda }{1-q}z\sum_{m\ge 0}\frac{(\lambda q,\lambda q^2;q^2)_{m}}{(q^3;q^2)_{m}(q^2;q^2)_m}(z^2)^{m}\\ 
&={}_2\varphi_1(\lambda ,\lambda q;q;q^2,z^2)+\frac{1-\lambda}{1-q}z {}_2\varphi_1(\lambda q,\lambda q^2;q^3;q^2,z^2)\\ 
&={}_2\varphi_1(\lambda ,\lambda q;q;q^2,z^2)+\frac{(\lambda ;q^2)_\infty}{(\lambda q^2;q^2)_\infty}\frac{(q^3;q^2)_\infty}{(q;q^2)_\infty}z {}_2\varphi_1(\lambda q,\lambda q^2;q^3;q^2,z^2)\\ 
&=\frac{(q^2;q^2)_\infty}{(q;q^2)_\infty}\left\{\frac{(q;q^2)_\infty}{(q^2;q^2)_\infty}{}_2\varphi_1(\lambda ,\lambda q;q;q^2,z^2)+\frac{(\lambda ,q^3;q^2)_\infty}{(\lambda q^2,q^2;q^2)_\infty}z {}_2\varphi_1(\lambda q,\lambda q^2;q^3;q^2,z^2)\right\}. 
\end{align*}
Therefore, we acquire the alternative representation of the $q$-binomial theorem
\begin{align}\label{qbin1}
{}_1\varphi_0(\lambda ;-;q,z)
=\frac{(q^2;q^2)_\infty}{(q;q^2)_\infty}&\left\{\frac{(q;q^2)_\infty}{(q^2;q^2)_\infty}{}_2\varphi_1(\lambda ,\lambda q;q;q^2,z^2)\right.\notag\\ 
&\hspace{-.5em} + \left. \frac{(\lambda ,q^3;q^2)_\infty}{(\lambda q^2,q^2;q^2)_\infty}z {}_2\varphi_1(\lambda q,\lambda q^2;q^3;q^2,z^2)\right\}. 
\end{align}
We now define the \textit{weight function}
\begin{equation}\label{weight}
w(z,\lambda ;q):=q^{+\frac{1}{8}}z^{-\frac{1}{2}}\frac{(z^2;q^2)_\infty}{(\lambda ;q^2)_\infty}
\end{equation}
to make the generalization of the relation \eqref{amirror3} clear. Multiplying the weight function \eqref{weight} by the alternative representation \eqref{qbin1}, we obtain
\begin{align}
w(z,\lambda ;q)\frac{(\lambda z;q)_\infty}{(z;q)_\infty}
&=q^{+\frac{1}{8}}z^{-\frac{1}{2}}\frac{(z^2;q^2)_\infty}{(\lambda ;q^2)_\infty}\frac{(\lambda z;q)_\infty}{(z;q)_\infty}\notag\\ 
&=q^{+\frac{1}{8}}z^{-\frac{1}{2}}\frac{(z^2;q^2)_\infty}{(\lambda ;q^2)_\infty}
 \frac{(q^2;q^2)_\infty}{(q;q^2)_\infty}\left\{\frac{(q;q^2)_\infty}{(q^2;q^2)_\infty}{}_2\varphi_1(\lambda ,\lambda q;q;q^2,z^2)\right.\notag\\ 
&\hspace{12.5em} + \left. \frac{(\lambda ,q^3;q^2)_\infty}{(\lambda q^2,q^2;q^2)_\infty}z {}_2\varphi_1(\lambda q,\lambda q^2;q^3;q^2,z^2)\right\}\notag\\ 
&=q^{+\frac{1}{8}}\frac{(q^2;q^2)_\infty}{(q;q^2)_\infty}
\left\{z^{-\frac{1}{2}}\frac{(z^2,q;q^2)_\infty}{(\lambda ,q^2;q^2)_\infty}{}_2\varphi_1(\lambda ,\lambda q;q;q^2,z^2)\right.\notag\\ 
&\hspace{6.75em} + \left. z^{+\frac{1}{2}}\frac{(z^2,q^3;q^2)_\infty}{(\lambda q^2,q^2;q^2)_\infty}{}_2\varphi_1(\lambda q,\lambda q^2;q^3;q^2,z^2)\right\}\notag, 
\end{align}
namely,
\begin{align}w(z,\lambda ;q)\frac{(\lambda z;q)_\infty}{(z;q)_\infty}&=q^{+\frac{1}{8}}\frac{(q^2;q^2)_\infty}{(q;q^2)_\infty}
\left\{z^{-\frac{1}{2}}\frac{(z^2,q;q^2)_\infty}{(\lambda ,q^2;q^2)_\infty}{}_2\varphi_1(\lambda ,\lambda q;q;q^2,z^2)\right.\notag\\ 
&\hspace{6.75em} + \left. z^{+\frac{1}{2}}\frac{(z^2,q^3;q^2)_\infty}{(\lambda q^2,q^2;q^2)_\infty}{}_2\varphi_1(\lambda q,\lambda q^2;q^3;q^2,z^2)\right\}. 
\label{kl1}
\end{align}

When we put $z\mapsto \tilde{a}^{+\frac{1}{2}}$ and $\lambda\mapsto \tilde{a}^{-1}q$, i.e. $\lambda z\mapsto \tilde{a}^{-\frac{1}{2}}q$ in \eqref{kl1}, we obtain the relation \eqref{amirror3}.

\subsection{From the triple product identity of the theta function of Jacobi} \label{theta}
Next, we prove the relation \eqref{amirror3} in terms of the theta function of Jacobi. The idea of the proof comes from the connection problems on linear $q$-difference equations \cite{Mbi}. The local theory and irregularity for $q$-difference equations are studied by J.-P.~Ramis, J.~Sauloy, and C.~Zhang \cite{RSZ} by using the Newton polygon. Recently, C.~Zhang and T.~Morita provided some connection formulas with the irregular singular case. In the connection problems, we study the elliptic functions associated with the relations between the local solutions around the origin and around infinity. In this subsection, we deal with the first-order $q$-difference equation (see Remark \ref{remark1} for details). We begin with the review of the theta function \cite{Mbi}. The theta function is given by 
\[\theta (x)=\sum_{n\in\mathbb{Z}}q^{\frac{n(n-1)}{2}}x^n, \quad \forall x\in\mathbb{C}.\]
The theta function has the triple product identity
\begin{equation}\label{triple}
\theta (x)=\lp q,-x,- \frac{q}{x} ;q \rp_\infty .
\end{equation}
For any $k\in\mathbb{Z}$, the theta function satisfies the $q$-difference equation
\begin{align}
\theta (q^kx)=q^{-\frac{k(k-1)}{2}}x^{-k}\theta (x).
\end{align}
The theta function also has the inversion formula
\begin{align}
\theta (1/x)=\theta (x)/x.
\end{align}

The function ${}_1\varphi_0(\lambda ;-;q,z)$ can be rewritten by using the theta function as
\begin{equation}\label{infpr1}
{}_1\varphi_0(\lambda ;-;q,z)=\frac{(\lambda z;q)_\infty}{(z;q)_\infty}
=\frac{\theta (-\lambda z)}{\theta (-z)}\frac{(q/z;q)_\infty}{(q/\lambda z;q)_\infty}=\frac{\theta (-\lambda z)}{\theta (-z)}
{}_1\varphi_0\left(\lambda ;-;q,\frac{q}{\lambda z}\right),
\end{equation}
provided that $|z|<1$.
\begin{rem}\label{remark1}
The function ${}_1\varphi_0(\lambda ;-;q,z)$ satisfies the first-order $q$-difference equation
\begin{equation}\label{qbeq}
(1-\lambda z)u(qz)+(z-1)u(z)=0.
\end{equation}
We can check that the equation \eqref{qbeq} has the solution around infinity
\begin{equation}\label{qsolinf}
u_\infty (z) :=\frac{\theta (\lambda z)}{\theta (z)}{}_1\varphi_0\left(\lambda ;-;q,\frac{q}{\lambda z}\right).
\end{equation}
With this solution, the relation \eqref{infpr1} can be rewritten as
\begin{align*}
{}_1\varphi_0\left(\lambda ;-;q,z\right)=C_q(z)u_\infty(z),
\end{align*}
where
\begin{align*}
C_q(z)=\frac{\theta (-\lambda z)}{\theta (-z)}\frac{\theta (z)}{\theta (\lambda z)}.
\end{align*}
Here, the function $C_q(z)$ is the elliptic function, namely, $q$-periodic and unique valued:
\[C_q(qz)=C_q(z),\quad C_q(e^{2\pi i}z)=C_q(z).\]
Therefore, the function $C_q(z)$ gives the ``true'' connection coefficient \cite{Mbi} between the function ${}_1\varphi_0(\lambda ;-;q,z)$ and $u_\infty (z)$.
\end{rem}

The function ${}_1\varphi_0(\lambda ;-;q,q/\lambda z)$ also has the \textit{alternative representation} \eqref{qbin1} as
\begin{align}
{}_1\varphi_0\left(\lambda ;-;q,\frac{q}{\lambda z}\right)
&={}_2\varphi_1\left(\lambda ,\lambda q;q;q^2,\left(\frac{q}{\lambda z}\right)^2\right)\notag\\ 
&\hspace{1em} +\frac{(\lambda ,q^3;q^2)_\infty}{(\lambda q^2,q;q^2)_\infty}\frac{q}{\lambda z}
{}_2\varphi_1\left(\lambda q,\lambda q^2;q^3;q^2,\left(\frac{q}{\lambda z}\right)^2\right).
\label{alt22}
\end{align}
Combining the relations \eqref{infpr1}, \eqref{alt22}, and the weight function $w(z,\lambda ;q)$ defined in Appendix \ref{qbin}, we also obtain the following relation:
\begin{align}
w(z,\lambda ;q)\frac{(\lambda z;q)_\infty}{(z;q)_\infty}
&=q^{+\frac{1}{8}}z^{-\frac{1}{2}}\frac{(z^2;q^2)_\infty}{(\lambda ;q^2)_\infty}\frac{(\lambda z;q)_\infty}{(z;q)_\infty}\notag\\ 
&=q^{+\frac{1}{8}}\frac{(q^2;q^2)_\infty}{(q;q^2)_\infty}\frac{\theta \left(-\frac{q}{\lambda z}\right)}{\theta\left(-z\right)}
\left\{z^{-\frac{1}{2}}\frac{(z^2,q;q^2)_\infty}{(\lambda ,q^2;q^2)_\infty}{}_2\varphi_1\left(\lambda ,\lambda q;q;q^2,\left(\frac{q}{\lambda z}\right)^2\right)\right.\notag\\ 
&\hspace{10.4em} + \left. z^{-\frac{1}{2}}\frac{q}{\lambda z}\frac{(z^2,q^3;q^2)_\infty}{(\lambda q^2,q^2;q^2)_\infty}{}_2\varphi_1\left(\lambda q,\lambda q^2;q^3;q^2,\left(\frac{q}{\lambda z}\right)^2\right)\right\}. 
\label{infinity}
\end{align}
Equation \eqref{infinity} gives the relation between the basic hypergeometric series ${}_1\varphi_0$ around the origin and the basic hypergeometric series ${}_2\varphi_1$ around infinity.

If we set $z\mapsto \tilde{a}^{+\frac{1}{2}}$ and $\lambda \mapsto \tilde{a}^{-1}q$, i.e. $\lambda z\mapsto \tilde{a}^{-\frac{1}{2}}q$, we again acquire the relation \eqref{amirror3}.

\providecommand{\href}[2]{#2}\begingroup\raggedright\endgroup
\end{document}